\documentclass[twocolumn, superscriptaddress,amsmath,amssymb,aps, pra,
showkeys,showpacs,nofootinbib] {revtex4-1}
\usepackage{graphicx}
\usepackage{color}
\usepackage{bm}
\usepackage{amsmath}
\usepackage{hyperref}
\usepackage{dcolumn}
\usepackage{slashed}
\usepackage{bm}

\begin{document}
\title{Concurrent Remote Entanglement with Quantum Error Correction}
\author{Ananda Roy}
\email{ananda.roy@yale.edu}
\affiliation{Department of Applied Physics, Yale University, PO BOX 208284, New Haven, CT 06511}
\author{A. Douglas Stone}
\affiliation{Department of Applied Physics, Yale University, PO BOX 208284, New Haven, CT 06511}
\author{Liang Jiang}
\affiliation{Department of Applied Physics, Yale University, PO BOX 208284, New Haven, CT 06511}

\begin{abstract}
Remote entanglement of distant, non-interacting quantum entities is a key primitive for quantum information processing. We present a new protocol to remotely entangle two stationary qubits by first entangling them with propagating ancilla qubits and then performing a joint two-qubit measurement on the ancillas. Subsequently, single-qubit measurements are performed on each of the ancillas. We describe two continuous variable implementations of the protocol using propagating microwave modes. The first implementation uses propagating Schr$\rm{\ddot{o}}$dinger cat-states as the flying ancilla qubits, a joint-photon-number-modulo-2 measurement of the propagating modes for the two-qubit measurement and homodyne detections as the final single-qubit measurements. The presence of inefficiencies in realistic quantum systems limit the success-rate of generating high fidelity Bell-states. This motivates us to propose a second continuous variable implementation, where we use quantum error correction to suppress the decoherence due to photon loss to first order. To that end, we encode the ancilla qubits in superpositions of Schr\"odinger cat states of a given photon-number-parity, use a joint-photon-number-modulo-4 measurement as the two-qubit measurement and homodyne detections as the final single-qubit measurements. We demonstrate the resilience of our quantum-error-correcting remote entanglement scheme to imperfections. Further, we describe a modification of our error-correcting scheme by incorporating additional individual photon-number-modulo-2 measurements of the ancilla modes to improve the success-rate of generating high-fidelity Bell-states. Our protocols can be straightforwardly implemented in state-of-the-art superconducting circuit-QED systems. 
\end{abstract}

\maketitle 
\section{Introduction}
\label{intro}
Generation of entangled states between spatially separated non-interacting quantum systems is crucial for large-scale quantum information processing. For instance, it is necessary for implementation of quantum cryptography using the Ekert protocol \cite{Ekert_1991}, teleportation of unknown quantum states \cite{Bennett_Wootters_1993} and efficient quantum communication over a distributed quantum network \cite{van_Enk_Zoller_1997, Briegel_Zoller_1998}. At the same time, it is also valuable for performing a loophole-free test of Bell's inequalities \cite{CHSH, Cabello_2001, Simon_Irvine_2003, Matsukevich_Monroe_2008}. In particular, a {\it concurrent} remote entanglement scheme, in which no signal propagates from one qubit to the other, is a desirable feature of a scalable, module-based architecture of quantum computing \cite{Duan_Monroe_2004, Jiang_Lukin_2007, Devoret_Schoelkopf_2013, Monroe_Kim_2014}. 

The inevitable presence of imperfections in current experimentally accessible quantum systems have stimulated a search for remote entanglement protocols that are resilient to these imperfections. Heralded remote entanglement schemes based on interference of single photons from distant excited atoms or atomic ensembles using beam-splitters and subsequent photon detection have been proposed \cite{Cabrillo_Zoller_1999, DLCZ, Barrett_Kok_2004} and demonstrated \cite{Chou_Kimble_2005, Moehring_Monroe_2007, Hofmann_Weinfurter_2012, Bernien_Hanson_2013}. These protocols make use of the inherent resilience of Fock states to photon loss arising out of imperfections. As a consequence, when a successful event happens, it leads to a very high fidelity entangled state. However, the collection and detection efficiencies limit the success probability of generating entangled states. Alternate protocols using continuous variables of microwave light, in particular superpositions of coherent states, have been proposed that have a high success rate \cite{Roy_Devoret_2015, Silveri_Girvin_2015}. However, in presence of imperfections, the success-rates of these protocols diminishes drastically for generating high fidelity entangled state. This is because superpositions of coherent states are extremely susceptible to decoherence due to photon loss. 
The goal of this paper is to propose a new, concurrent, continuous-variable, remote entanglement protocol, which is amenable to quantum error correction to suppress the decoherence due to photon loss. 

The protocol can be summarized as follows.
In order to generate entanglement between two distant, stationary qubits, we use a propagating ancilla qubit for each of the stationary qubits. In the first step, each of the stationary qubits is entangled with its associated propagating ancilla qubit. In the next step, a two-qubit measurement (ZZ) is performed on the propagating ancillas. This non-linear measurement erases the `which stationary qubit is entangled to which flying qubit information' and gives rise to four-qubit entangled states. The final step comprises of a single qubit measurement on each ancilla qubit, to disentangle them from the stationary qubits, and finally prepare the desired entangled states between the two stationary qubits.

We describe two continuous-variable implementations of the aforementioned protocol. In the first implementation, we encode the ancilla qubits in Schr\"odinger cat states of propagating modes of microwave light \cite{Jeong_Kim_2002, Ralph_Milburn_2002, Leghtas_Mirrahimi_2013, Mirrahimi_Devoret_2014}. The logical basis states of each of the ancilla qubits are mapped to even and odd Schr\"{o}dinger cat states, denoted by $|C_{\alpha}^{\pm}\rangle$, defined below: 
\begin{equation}
\label{two_cat_defn}
|C_{\alpha}^{\pm}\rangle = {\cal N}_{\pm}\big(|\alpha\rangle\pm|-\alpha\rangle\big),
\end{equation}
where ${\cal N}_\pm = 1/\sqrt{2(1\pm e^{-2|\alpha|^2})}$. 
The two-qubit measurement (ZZ) on the ancillas is a joint-photon-number-modulo-$2$ measurement, while the single-qubit measurements are homodyne detections. In absence of imperfections, this protocol gives rise to maximally entangled Bell-states with unit probability. 
However, in presence of imperfections, photon loss
lead to decoherence of the propagating (ancilla qubit) microwave modes which are entangled with the stationary qubits. 
This limits the success rate of generating high fidelity Bell-states. 

To remedy this effect, we propose a second implementation of our protocol. In this implementation, we use a different encoding of the ancilla qubits, where the logical basis states are mapped to the states $|C_{\alpha}^{0,2{\rm{mod}}4}\rangle$, hereafter referred to as ``mod 4 cat states" \cite{Leghtas_Mirrahimi_2013, Mirrahimi_Devoret_2014, Ofek_Schoelkopf_2016}, of a propagating temporal mode, defined below:
\begin{eqnarray}
\label{four_cat_defn}
\big|C_{\alpha}^{0{\rm{mod}}4}\big\rangle &=& \frac{1}{\sqrt{2\big\{1+\frac{\cos(|\alpha|^2)}{\cosh(|\alpha|^2)}\big\}}}\big(\big|C_{\alpha}^{+}\big\rangle+\big|C_{i\alpha}^{+}\big\rangle\big),\nonumber\\\big|C_{\alpha}^{2{\rm{mod}}4}\big\rangle &=& \frac{1}{\sqrt{2\big\{1-\frac{\cos(|\alpha|^2)}{\cosh(|\alpha|^2)}\big\}}}\big(\big|C_{\alpha}^{+}\big\rangle-\big|C_{i\alpha}^{+}\big\rangle\big).
\end{eqnarray}
The state $|C_{\alpha}^{0(2){\rm{mod}}4}\rangle$ has photon-number populations in the Fock states $4n(4n+2), n\in\mathbb{N}$, which is indicated by the notation $0(2) \rm{mod} 4$. For this encoding, the two-qubit ZZ measurement is a joint-photon-number-modulo-4 measurement, while the single-qubit measurements are homodyne detections.
In absence of imperfections, the joint-photon-number-modulo-4 outcome can be either 0 or 2. Now consider the case when there are imperfections. Photon loss due to these imperfections takes the populations of the propagating temporal mode from the even photon-number-parity manifold to the odd-photon-number-parity manifold. 
This change in photon-number-parity changes the outcome of the joint-photon-modulo-4 measurement. By detecting this change of the joint-photon-number-modulo-4 measurement outcome, we correct for the decoherence of the entangled qubit-photon states due to loss of a photon in either of the ancillas. Furthermore, additional individual photon-number-modulo-2 measurements of the ancillas, in addition to the joint-photon-number-modulo-4 measurement, suppress the loss of coherence due to loss of a single-photon in both ancillas. 


Superconducting circuit-QED systems have access to a tunable, strong and dispersive nonlinearity, in the form of the Josephson nonlinearity. Moreover, the collection and detection efficiency of microwave photons are significantly better than their optical counterparts. These advantages make the superconducting circuit-QED systems natural platforms for implementing these protocols. In fact, sequential interaction of a propagating microwave photon mode with two distant qubits, using linear signal processing techniques, have successfully entangled the two qubits \cite{Nemoto_Munro_2004, Spiller_MIlburn_2006, Roch_Siddiqi_2014, Sarlette_Mirrahimi_2016}. Furthermore, measurement of joint-photon-number-modulo-2 of two cavity modes has already been demonstrated in circuit-QED systems \cite{Wang_Schoelkopf_2016}. 

The paper is organized as follows. Sec. \ref{prot_flying_qubits} describes our protocol using propagating ancilla qubits. Sec. \ref{cont_var_impl} describes two continuous variable implementations of this protocol for the two different encodings of the ancilla qubits mentioned above. Secs. \ref{mod 2_imperf}, \ref{mod 4_imperf} incorporate the effect of imperfections like undesired photon loss and detector inefficiencies for the two implementations. Sec. \ref{mod 2_mod 4_imperf_comp} compares the resilience of the two implementations to these imperfections. Sec. \ref{mod 4_p1p2} discusses the improvement to our protocol by incorporating additional, individual, photon-number-modulo-2 measurements. Our results are summarized and future directions are outlined in Sec. \ref{concl}. 
\section{Protocol using propagating ancilla qubits}
\label{prot_flying_qubits}
\begin{figure}
\centering
\includegraphics[width = 0.5\textwidth]{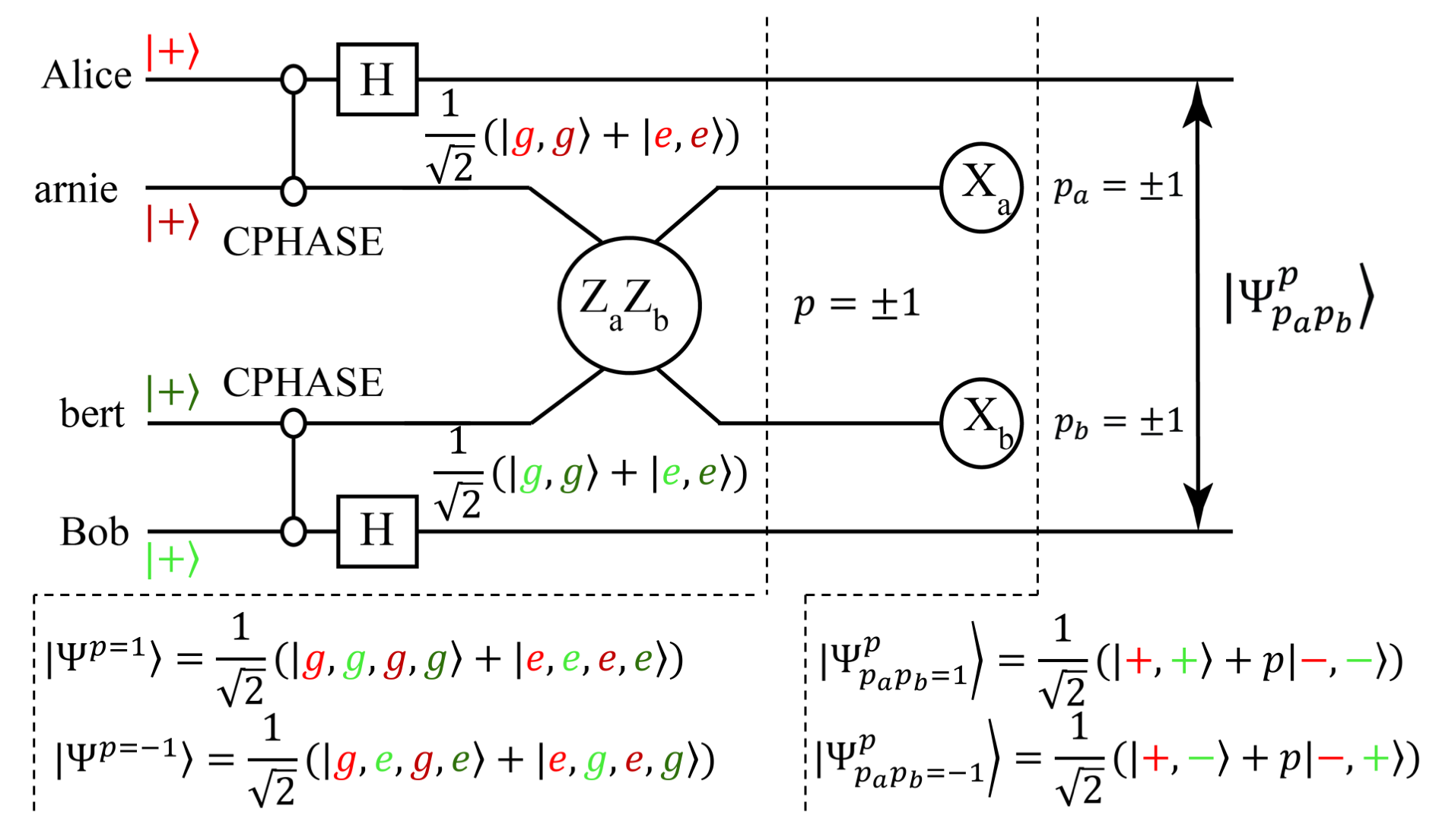}
\caption{ \label{fig_1} (color online) Remote entanglement protocol schematic. The first step of the protocol comprises of local entanglement generation between two stationary, mutually non-interacting qubits, Alice (in red) and Bob (in green), with propagating ancilla qubits, arnie (in dark red) and bert (in dark green). To that end, the four qubits are initialized to their respective $|+\rangle$ states. Subsequently, a CPHASE gate is applied between Alice (Bob) and arnie (bert), followed by a Hadamard rotation on Alice (Bob). After this step, the entangled state of Alice (Bob) and arnie (bert) is $(|g,g\rangle + |e,e\rangle)/\sqrt{2}$. Next, a two-qubit measurement, $Z_aZ_b$, is performed on arnie and bert. Conditioned on the measurement outcome $p=\pm1$, a four-qubit entangled state is generated: $|\Psi^{p=1}\rangle = (|g,g,g,g\rangle + |e,e,e,e\rangle)/\sqrt{2}$ or $|\Psi^{p=-1}\rangle = (|g,e,g,e\rangle + |e,g,e,g\rangle)/\sqrt{2}$. Subsequently, single-qubit measurements (X) are performed on arnie and bert, denoted by $X_a, X_b$, with measurement outcomes $p_a, p_b=\pm1$. Conditioned on the three measurement outcomes $p, p_a, p_b$, Alice and Bob are projected onto a paritcular entangled state $|\Psi^p_{p_ap_b=1}\rangle = (|+,+\rangle + p|-,-\rangle)/\sqrt{2}$ or $|\Psi^p_{p_ap_b=-1}\rangle = (|+,-\rangle + p|-,+\rangle)/\sqrt{2}$.}
\end{figure}

In this section, we present a concise description of our protocol to entangle two stationary, mutually non-interacting qubits, Alice (A) and Bob (B), using two propagating ancilla qubits, arnie (a) and bert (b) (cf. Fig.\  \ref{fig_1}). Each qubit is initialized to its $|+\rangle=(|g\rangle+|e\rangle)/\sqrt{2}$ state. Local entanglement is generated between Alice (Bob) and arnie (bert), by first applying a CPHASE gate between Alice (Bob) and arnie (bert), followed by a Hadamard gate on Alice (Bob). After this step, the state of Alice (Bob) and arnie (bert) is: $(|g,g\rangle + |e,e\rangle)/\sqrt{2}$. Subsequently, a quantum non-demolition two-qubit measurement, $Z_aZ_b$,  is performed on arnie and bert, whose outcome is denoted by $p = \pm1$. This measurement gives rise to one of the two four-qubit entangled states: $|\Psi^{p=1}\rangle = (|g,g,g,g\rangle + |e,e,e,e\rangle)/\sqrt{2}$ and  $|\Psi^{p=-1}\rangle = (|g,e,g,e\rangle + |e,g,e,g\rangle)/\sqrt{2}$. Here, the first, second, third and fourth position in the kets belong to the states of Alice, Bob, arnie and bert respectively. Since the final aim is to generate an entangled state of just Alice and Bob, arnie and bert must be disentangled from Alice and Bob, while preserving the entanglement between the latter two. This is the task of the final step. It comprises of making X-measurements on arnie and bert, whose outcomes are denoted by $p_a(=\pm1)$ and $p_b(=\pm1)$ respectively. Conditioned on the three measurement outcomes $p,p_a, p_b$, Alice and Bob get entangled with each other, with the final state being: $|\Psi^p_{p_ap_b =1}\rangle = (|+,+\rangle + p|-,-\rangle)/\sqrt{2}$ or $|\Psi^p_{p_ap_b=-1}\rangle = (|+,-\rangle + p|-,+\rangle)/\sqrt{2}$. 

In what follows, we describe proposals for the continuous variable implementations that realize the aforementioned protocol.
\section{Implementation Using Propagating Superpositions of Coherent States}
\label{cont_var_impl}
\begin{figure}
\centering
\includegraphics[width = 0.5\textwidth]{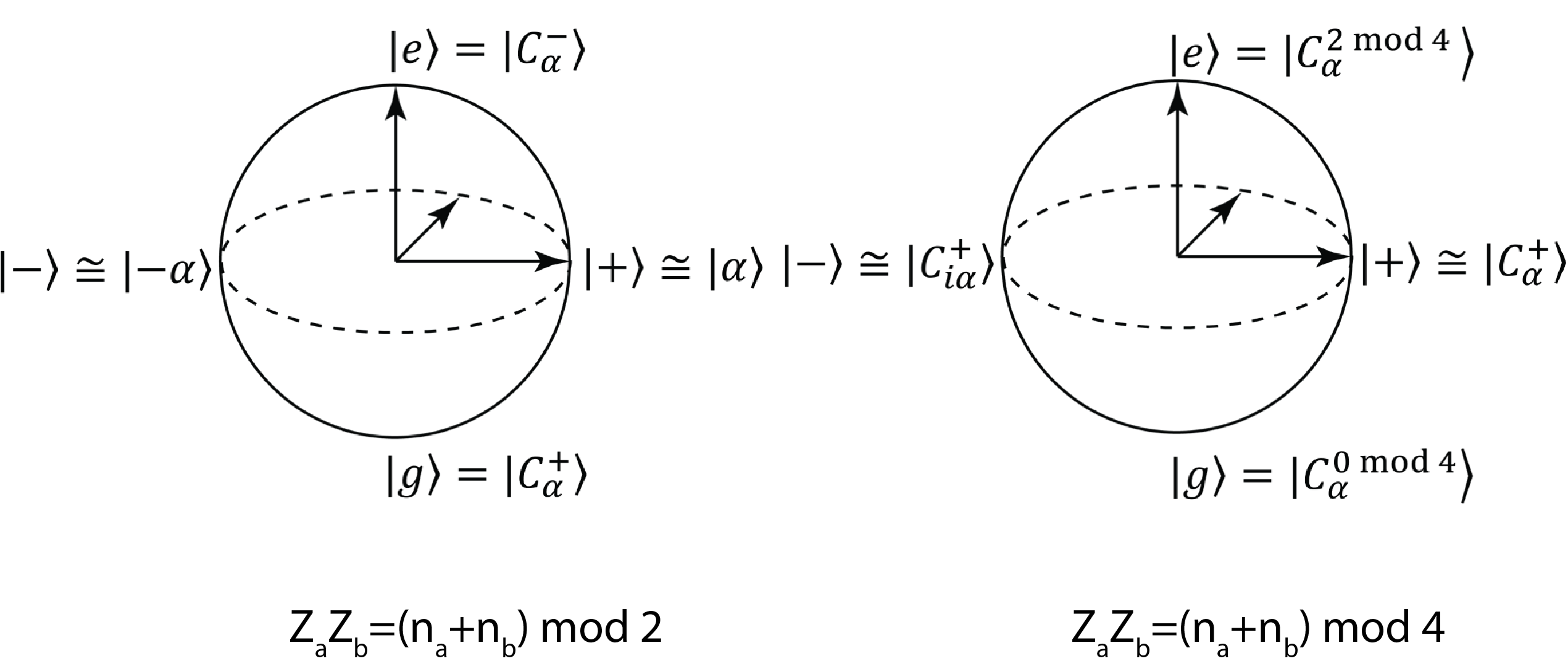}
\caption{ \label{fig_2} (color online) Cat-qubit mapping schematic. (Left) The ground (excited) state of each of the ancilla qubits is mapped to even (odd) Schr\"odinger cat states $|C_{\alpha}^{+(-)}\rangle$ [see Eq.\ \eqref{two_cat_defn}]. Consequently, the states $|\pm\rangle$ are mapped to coherent states $|\pm\alpha\rangle$. In this mapping, $Z_aZ_b$ on the propagating ancilla qubits corresponds to a joint photon-number-modulo-$2$ measurement of the propagating microwave modes. (Right) The ground (excited) state of each of the ancilla qubits is mapped to mod 4 cat states $|C_{\alpha}^{0(2){\rm{mod}}4}\rangle$ [see Eq.\ \eqref{four_cat_defn}]. Consequently, $|\pm\rangle$ are mapped to even cat states $|C_{\alpha(i\alpha)}^+\rangle$. In this mapping, $Z_aZ_b$ on the propagating ancilla qubits corresponds to a  joint photon-number-modulo-$4$ measurement of the propagating microwave modes. For both encodings, the single-qubit measurements (X) can be implemented by homodyne detections (not shown in the figure for brevity).}
\end{figure}

The first implementation uses the mapping of the ground (excited) state of the ancilla qubits to even (odd) Schr\"odinger cat states  $|C_{\alpha}^{+(-)}\rangle$ (cf. Fig.\ \ref{fig_2}). Consequently, the states $|\pm\rangle$ are approximately mapped to coherent states $|\pm\alpha\rangle$. The ZZ measurement is performed by a joint-photon-number-modulo-2 measurement, while the X measurements are performed by homodyne detections. In the second implementation, we map the ground (excited) states of the ancillas to the mod 4 cat states $|C_{\alpha}^{0(2){\rm{mod}}4}\rangle$, whence the $|+(-)\rangle$ are approximately mapped to even cat states $|C_{\alpha(i\alpha)}^+\rangle$. In this case, the ZZ measurement is achieved by a joint-photon-number-modulo-4 measurement, while the X measurements are performed by homodyne detections. In this section, we treat the case of perfect quantum efficiency, and incorporate imperfections in our computational model in Sec. \ref{FinEff}.

\subsection{Implementation using Schr\"odinger cat states}
\label{mod 2_perf}
In the first step of the protocol, local entanglement is generated between a stationary transmon qubit, Alice (Bob), and a propagating microwave mode, arnie (bert), giving rise to the following states: $(|g,C_\alpha^+\rangle+|e,C_\alpha^-\rangle)/\sqrt{2}$ ($(|g,C_\beta^+\rangle+|e,C_\beta^-\rangle)/\sqrt{2}$). This specific entangled state can be generated by first generating this entangled state inside a qubit-cavity system using the protocol proposed in \cite{Krastanov_Jiang_2015} and experimentally demonstrated in \cite{Heeres_Schoelkopf_2015}. Without loss of generality, we choose $\alpha,\beta\in\Re, \alpha=\beta>0$. 
We require the temporal profile of the modes of arnie and bert as they fly away from Alice and Bob to be $e^{\kappa_a t/2}\cos(\omega_a t)\Theta(-t)$  and $e^{\kappa_b t/2}\cos(\omega_b t)\Theta(-t)$ respectively, where $\omega_{a,b}, \kappa_{a,b}$ are defined below. The specific temporal mode profile can be implemented using a Q-switch \cite{Cirac_Mabuchi_1997, Yin_Martinis_2013} and is necessary for these modes to be subsequently captured in resonators for the joint-photon-number-modulo-2 measurement. The total state of the system, comprising of Alice, Bob, arnie and bert, can be written as: 
\begin{eqnarray}
\label{psi_even_odd_step_1_eqn}
|\Psi_{\rm{ABab}}\rangle &=& \frac{1}{2}\big(|g,g,C_\alpha^+,C_\alpha^+\rangle + |e,e,C_\alpha^-,C_\alpha^-\rangle\nonumber\\&& + |g,e,C_\alpha^+,C_\alpha^-\rangle + |e,g,C_\alpha^-,C_\alpha^+\rangle\big)
\end{eqnarray}

Next, a joint-photon-number-modulo-$2$ measurement is performed on these propagating microwave modes as follows.
The propagating microwave modes, entangled with the stationary qubits, pass through transmission lines and are resonantly incident on two cavities, exciting their fundamental modes with frequencies (decay rates) $\omega_a (\kappa_a)$ and $\omega_b (\kappa_b)$, respectively. Due to the specific form of the chosen mode-profile, at $t=0$, these propagating modes get perfectly captured in these cavities. Subsequently, their joint-photon-number-modulo-$2$ is measured by coupling a transmon qubit to these modes \cite{Wang_Schoelkopf_2016}. An even (odd) joint-photon-number-modulo-$2$ outcome corresponds to a measurement result $p=+1$($p=-1$) and the four-mode state, in absence of transmission losses and measurement imperfections, can be written as: 
\begin{eqnarray}
\label{p=1_eqn}
|\Psi^{p=1}_{\rm{ABab}}\rangle &=& \frac{1}{\sqrt{2}}\big(|g,g,C_\alpha^+,C_\alpha^+\rangle + |e,e,C_\alpha^-,C_\alpha^-\rangle\big),\\\label{p=-1_eqn}|\Psi^{p=-1}_{\rm{ABab}}\rangle &=& \frac{1}{\sqrt{2}}\big(|g,e,C_\alpha^+,C_\alpha^-\rangle + |e,g,C_\alpha^-,C_\alpha^+\rangle\big).
\end{eqnarray}

After this measurement, the ancilla qubits, arnie and bert, are entangled to Alice and Bob. The last step of the protocol performs the crucial function of disentangling Alice and Bob from the propagating ancillas, while preserving the entanglement between Alice and Bob. This is done by performing homodyne measurement along the direction ${\rm{arg}}(\alpha)$ of each of the outgoing microwave modes. Since we have chosen $\alpha\in\Re$, the X-quadratures of the microwave modes need to be measured. 
From Eqs.\ \eqref{p=1_eqn}, \eqref{p=-1_eqn}, it is evident that the pair of outcomes of the integrated homodyne signal $(x_a, x_b)$ in the vicinity of $(\alpha,\alpha)$ or $(-\alpha,-\alpha)$ projects Alice and Bob to $(|+,+\rangle + p|-,-\rangle)/\sqrt{2}$. Similarly, an outcome in the vicinity of  $(\alpha,-\alpha)$ or $(-\alpha,\alpha)$ projects Alice and Bob to $(|+,-\rangle + p|-,+\rangle)/\sqrt{2}$. 

For a given $p$, after the homodyne detections of arnie and bert, the density matrix $\rho^p_{\rm{ABab}} = |\Psi^p_{\rm{ABab}}\rangle\langle\Psi^p_{\rm{ABab}}|$ evolves to: 
\begin{equation}
\rho^p_{\rm{ABab}} \rightarrow \frac{{\cal M}_{X}\rho^p_{\rm{ABab}} {\cal M}_X^\dagger}{{\rm Tr}\big[{\cal M}_{X}\rho^p_{\rm{ABab}} {\cal M}_X^\dagger\big]},\hspace{0.1cm} {\cal M}_X = |x_a,x_b\rangle\langle x_a, x_b|.
\end{equation}
The probability distribution of the outcomes $P^p(x_a, x_b) = \frac{1}{2}{\rm Tr}\big[{\cal M}_{X}\rho^p_{\rm{ABab}} {\cal M}_X^\dagger\big]$. The factor of 1/2 arises because each of the outcomes $p=\pm1$ occurs with 1/2 probability. The resulting state of Alice and Bob is obtained by tracing out the states of arnie and bert. We evaluate $P^p(x_a, x_b)$ and the corresponding overlap to the Bell-states $|\phi^\pm\rangle = (|g,g\rangle \pm |e,e\rangle)/\sqrt{2}$  and $|\psi^\pm\rangle = (|g,e\rangle \pm |e,g\rangle)/\sqrt{2}$ to be (see Appendix \ref{appAmod2}):
\begin{eqnarray}
\label{prob_dist_mod2_perf}
P^{p}(x_a,x_b) &=& \frac{2}{\pi}e^{-2(x_a^2 + x_b^2)}e^{-4\alpha^2}N_p,\\
\label{overlap_mod2_perf}\langle\phi^\pm|\rho^p_{\rm{AB}}|\phi^{\pm}\rangle&=&\frac{1+p}{2}\Big[\frac{1}{2}\pm\frac{\sinh(4x_a\alpha)\sinh(4x_b\alpha)}{4N_p(1-e^{-4\alpha^2})}\Big],\\\label{overlap_mod2_perf_1}\langle\psi^\pm|\rho^p_{\rm{AB}}|\psi^{\pm}\rangle&=&\frac{1-p}{2}\Big[\frac{1}{2}\pm\frac{\sinh(4x_a\alpha)\sinh(4x_b\alpha)}{4N_p(1-e^{-4\alpha^2})}\Big],\ 
\end{eqnarray}
where \begin{eqnarray}
N_{p=1} &=& \Big[\frac{\cosh^2(2x_a\alpha)\cosh^2(2x_b\alpha)}{(1+e^{-2\alpha^2})^2}\nonumber\\&&+\frac{\sinh^2(2x_a\alpha)\sinh^2(2x_b\alpha)}{(1-e^{-2\alpha^2})^2}\Big],\\N_{p=-1} &=& \frac{1}{1-e^{-4\alpha^2}}\Big[\cosh^2(2x_a\alpha)\sinh^2(2x_b\alpha)\nonumber\\&&+\sinh^2(2x_a\alpha)\cosh^2(2x_b\alpha)\Big].
\end{eqnarray}
\begin{figure}
\centering
\includegraphics[width = 0.5\textwidth]{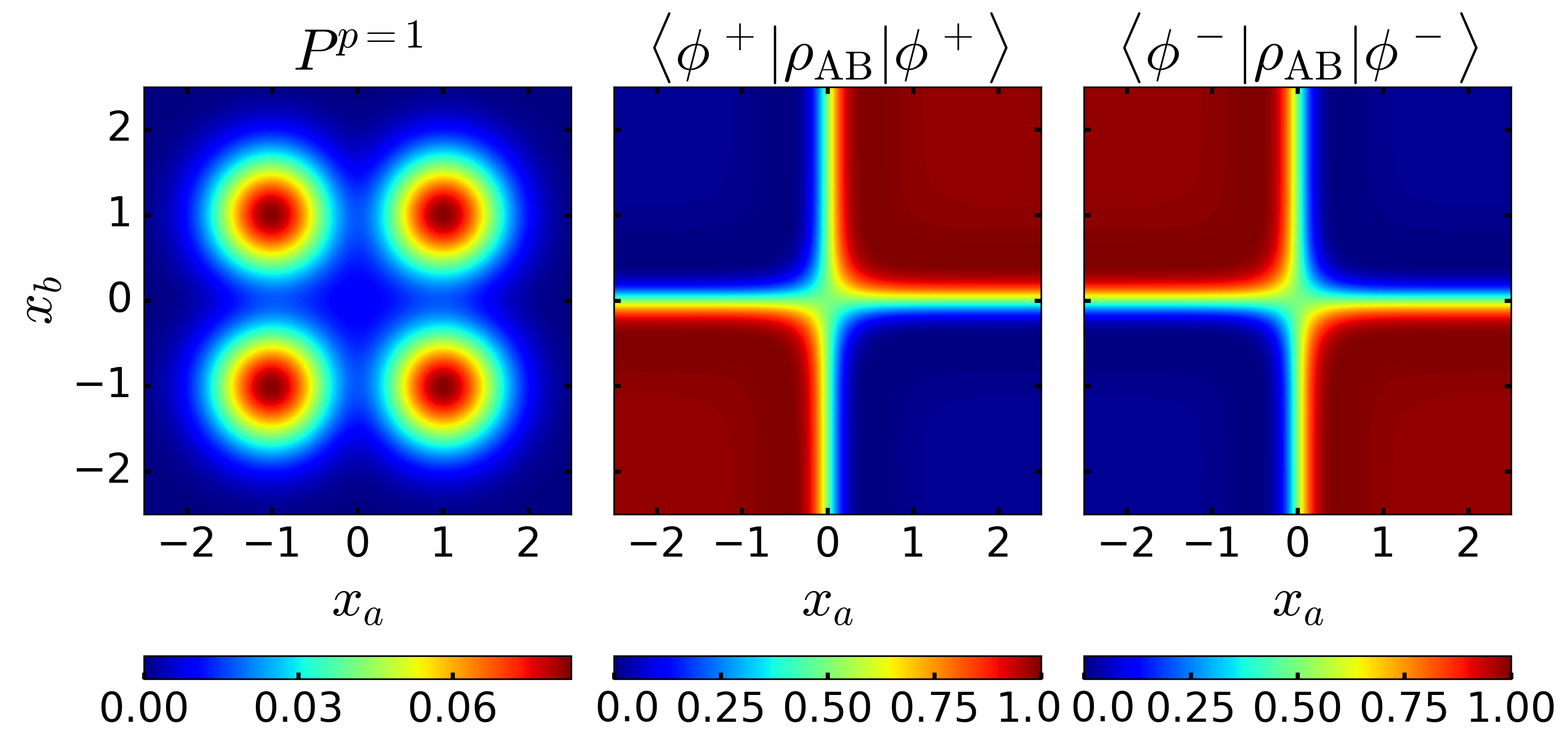}
\caption{ \label{fig_3} (color online) Probability distribution $P^p(x_a, x_b)$ of outcomes of the homodyne measurements of arnie and bert and resulting overlaps of Alice and Bob's joint density matrix $\rho_{\rm{AB}}$ to the Bell-states $|\phi^\pm\rangle = (|g,g\rangle \pm|e,e\rangle)/\sqrt{2}$ are shown for the case when the joint-photon-number-modulo-2 measurement yields $p=1$. We choose $\alpha=1$ and assume absence of measurement imperfections and photon loss. (Left) Probability distribution showing four Gaussian distributions centered at $x_a = \pm\alpha, x_b = \pm\alpha$. (Center and Right) Corresponding overlap to the Bell-state $|\phi^+\rangle$ tends to $1$ for $(x_a, x_b)$ in the vicinity of $(\alpha,\alpha)$ and $(-\alpha,-\alpha)$. Similarly, overlap to the Bell-state $|\phi^-\rangle$ tends to $1$ for $(x_a, x_b)$ in the vicinity of $(-\alpha,\alpha)$ and $(\alpha,-\alpha)$. For an outcome on one of the lines: $x_a=0$ or $x_b=0$, the resultant state of Alice and Bob is an equal superposition of $|\phi^+\rangle$ and $|\phi^-\rangle$ and is not an entangled state. For $p=-1$, identical results are obtained with the substitution: $|\phi^\pm\rangle\rightarrow|\psi^\pm\rangle$.}
\end{figure}

Fig.\ \ref{fig_3} shows the probability distribution $P^p(x_a, x_b)$ of the outcomes of the integrated homodyne currents $x_a, x_b$, together with the overlap to the Bell-states $|\phi^+\rangle, |\phi^-\rangle$ for the case when the joint-photon-number-modulo-2-measurement outcome is $p=1$. We choose $\alpha = 1$ in absence of transmission loss and measurement inefficiency. The probability distribution [Eq.\ \eqref{prob_dist_mod2_perf}] contains four Gaussian distributions centered around $x_a=\pm\alpha, x_b=\pm\alpha$. For $(x_a, x_b)$ in the vicinity of $(\alpha,\alpha)$ and $(-\alpha,-\alpha)$, the overlap to the Bell-state $|\phi^+\rangle$ approaches 1, while  for $(x_a, x_b)$ in the vicinity of $(\alpha,-\alpha)$ and $(-\alpha,\alpha)$, the overlap to the Bell-state $|\phi^-\rangle$ approaches 1. For outcomes along the lines $x_a =0$ and $x_b=0$, Alice and Bob are projected on to equal superpositions of $|\phi^+\rangle$ and $|\phi^-\rangle$ and thus, are not entangled. 
The results for the case $p=-1$ are identical with $|\phi^\pm\rangle$ is replaced by $|\psi^\pm\rangle$.

\subsection{Implementation using mod 4 cat states}
\label{mod 4_perf}
\begin{figure}
\centering
\includegraphics[width = 0.5\textwidth]{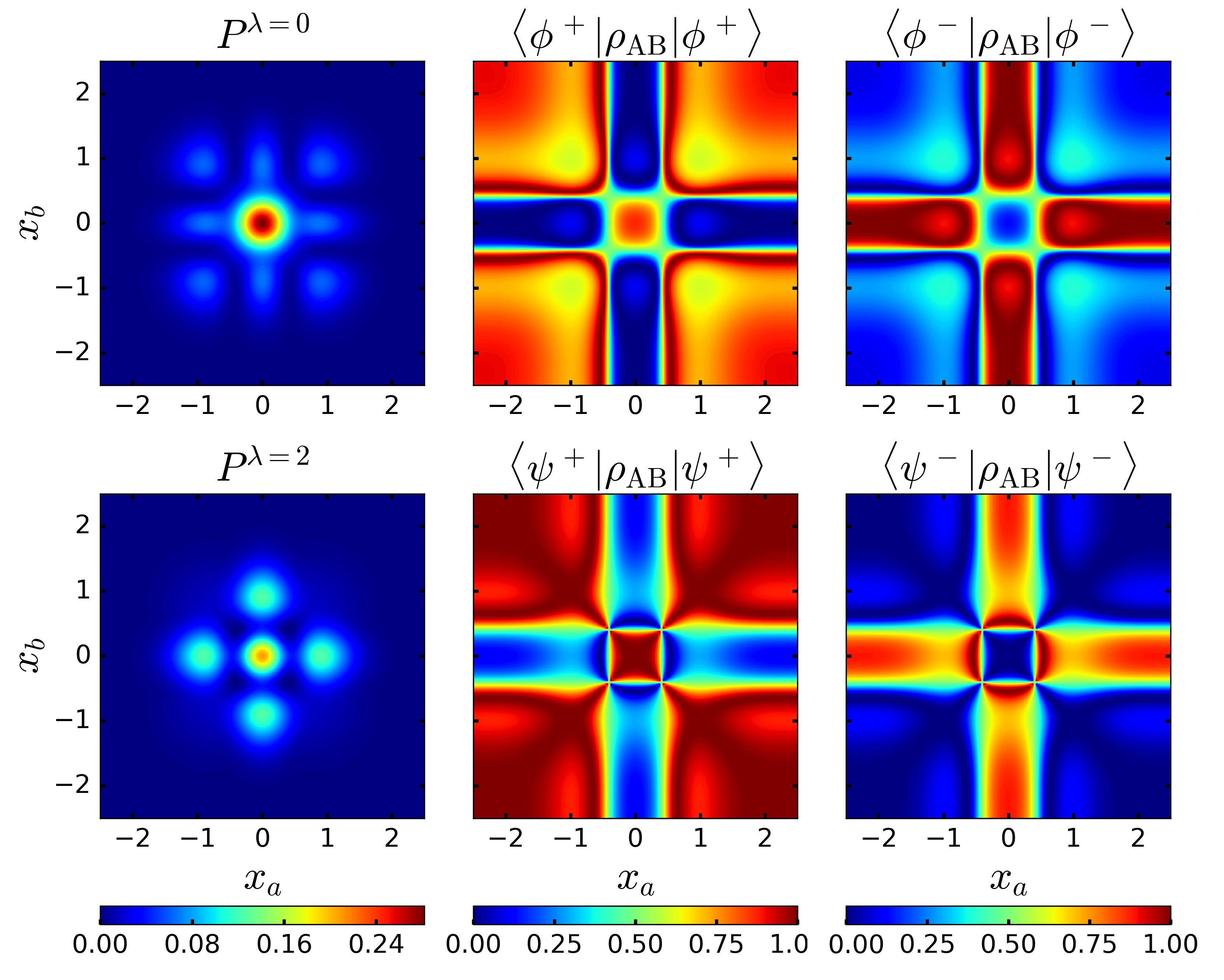}
\caption{ \label{fig_4} (color online) Probability distribution $P^\lambda(x_a, x_b)$ of outcomes of the homodyne measurements of arnie and bert and resulting overlaps of Alice and Bob's joint density matrix $\rho_{\rm{AB}}$ with the Bell-states $|\phi^\pm\rangle = (|g,g\rangle \pm|e,e\rangle)/\sqrt{2}, |\psi^\pm\rangle = (|g,e\rangle \pm|e,g\rangle)/\sqrt{2}$ are shown. We choose $\alpha=1$ in absence of measurement imperfections and photon loss. The top (bottom) left panel shows the probability of outcomes for the joint-photon-number-modulo-4 outcome $\lambda=0(2)$. Corresponding overlaps to the Bell-states $|\phi^\pm\rangle(|\psi^\pm\rangle)$ are plotted in the top (bottom) center and top (bottom) right panels. The overlaps to the odd (even) Bell-states for $\lambda=0(2)$ are zero and not shown for brevity. For both $\lambda=0$ and $\lambda=2$, one gets entangled Bell-states for Alice and Bob for majority of outcomes in the $(x_a,x_b)$ space. The alternating bright and dark fringes in plots are due to the measurement of X-quadrature of both arnie and bert, both of which are in superpositions of $|C_\alpha^+\rangle$ and $|C_{i\alpha}^+\rangle$. The size of the fringes decreases with increasing values of $\alpha$. }
\end{figure}

In this section, we describe the implementation of our protocol where the ancilla qubits are mapped on to mod 4 cat states. 
The first step of the protocol again involves generating entanglement between the stationary qubit of  Alice (Bob) and the propagating microwave mode arnie (bert) giving rise to the following states: $(|g,C_\alpha^{0{\rm{mod}}4}\rangle+|e,C_\alpha^{2{\rm{mod}}4}\rangle)/\sqrt{2}$ ($(|g,C_\beta^{0{\rm{mod}}4}\rangle+|e,C_\beta^{2{\rm{mod}}4}\rangle)/\sqrt{2}$). This set of entangled states can be obtained in an analogous manner using the method described in the previous subsection. 
We will again choose, without loss of generality, $\alpha,\beta\in\Re, \alpha=\beta>0$. 
The total state of the system, comprising of Alice, Bob, arnie and bert, can be written as: 
\begin{eqnarray}
\label{eqmod4}
|\Psi_{\rm{ABab}}\rangle &=& \frac{1}{2}\big(|g,g,C_\alpha^{0{\rm{mod}}4},C_\alpha^{0{\rm{mod}}4}\rangle\nonumber\\&& + |e,e,C_\alpha^{2{\rm{mod}}4},C_\alpha^{2{\rm{mod}}4}\rangle + |g,e,C_\alpha^{0{\rm{mod}}4},C_\alpha^{2{\rm{mod}}4}\rangle\nonumber\\&& + |e,g,C_\alpha^{2{\rm{mod}}4},C_\alpha^{0{\rm{mod}}4}\rangle\big)\nonumber.
\end{eqnarray}
By suitably engineering the temporal mode profiles of the propagating modes as in the previous subsection, these propagating entangled qubit-photon states are then captured in resonators. Subsequently, a joint-photon-number-modulo-4 measurement is performed on these captured modes (see Appendix. \ref{mod 4-impl} for details of the measurement protocol). In absence of measurement imperfections and losses, the joint-photon-number-modulo-4 has two possible outcomes: $\lambda = 0,2$ (the two-qubit measurement outcome $p$ of Sec. \ref{prot_flying_qubits} can be written as $p=i^\lambda$), corresponding to which the four-mode state can be written as:  
\begin{eqnarray}
\label{lambda=0_eqn}
|\Psi^{\lambda=0}_{\rm{ABab}}\rangle &=& \frac{1}{\sqrt{2}}\big(|g,g,C_\alpha^{0{\rm{mod}}4},C_\alpha^{0{\rm{mod}}4}\rangle\nonumber\\&& + |e,e,C_\alpha^{2{\rm{mod}}4},C_\alpha^{2{\rm{mod}}4}\rangle\big),\\\label{lambda=2_eqn}|\Psi^{\lambda=2}_{\rm{ABab}}\rangle &=& \frac{1}{\sqrt{2}}\big(|g,e,C_\alpha^{0{\rm{mod}}4},C_\alpha^{2{\rm{mod}}4}\rangle\nonumber\\&& + |e,g,C_\alpha^{2{\rm{mod}}4},C_\alpha^{0{\rm{mod}}4}\rangle\big).
\end{eqnarray}
The final step of the protocol comprises of making homodyne detections of arnie and bert and here we choose the X-quadrature of both these modes. Similar calculations can be done for other choices. Consider the case when $\lambda = 0$. From Eq. \eqref{lambda=0_eqn}, it follows that each homodyne detector will have Gaussian distributions centered around $x_a,x_b=0,\pm\alpha$. It also follows from Eq. \eqref{lambda=0_eqn} that for events $(x_a,x_b)$ in the five vicinity regions of $(\pm\alpha,\pm\alpha)$ and $(0,0)$, the resulting state of Alice and Bob is $|\phi^+\rangle$, while for outcomes in the vicinity of $(0,\pm\alpha)$ and $(\pm\alpha,0)$, the resulting state of Alice and Bob is $|\phi^-\rangle$. Similar set of reasoning holds for $\lambda=2$, when the states $|\psi^\pm\rangle$ are generated. Since the state of Alice and Bob depend only on ($|x_a|, |x_b|$), the resulting overlap distributions respect a four-fold rotational symmetry in the $(x_a,x_b)$ space (see Fig. \ref{fig_4}). 

After the homodyne detection, the density matrix of Alice, Bob, arnie and bert evolves to: 
\begin{equation}
\rho^\lambda_{\rm{ABab}} \rightarrow \frac{{\cal M}_{X}\rho^\lambda_{\rm{ABab}} {\cal M}_X^\dagger}{{\rm Tr}\big[{\cal M}_{X}\rho^\lambda_{\rm{ABab}} {\cal M}_X^\dagger\big]},\ {\cal M}_X = |x_a,x_b\rangle\langle x_a, x_b|. 
\end{equation}
The probability distribution of the outcomes $P^\lambda(x_a, x_b) = \frac{1}{2}{\rm Tr}\big[{\cal M}_{X}\rho^\lambda {\cal M}_X^\dagger\big]$. Subsequent state of Alice and Bob can again be obtained by tracing out the modes arnie and bert. Note again the factor of $1/2$ in the expression for probability distribution due to the probability 1/2 of occurrence of  either $\lambda=0$ or $\lambda=2$.

Computing the probability of outcomes and the overlap to the Bell-states (see Appendix. \ref{appAmod4} for details), we arrive at: 
\begin{eqnarray}
\label{prob_dist_mod4_perf}
P^\lambda(x_a,x_b) &=& \frac{1}{2\pi}\frac{e^{-2(x_a^2+x_b^2)}}{(1+e^{-2\alpha^2})^2}\tilde{N}_\lambda,\\\label{overlap_mod4_perf}\langle\phi^\pm|\rho^\lambda_{\rm{AB}}|\phi^{\pm}\rangle&=&\frac{1+i^\lambda}{2}\bigg[\frac{1}{2}\pm\frac{\prod\limits_{\delta=0,2}F_\delta(x_a)F_\delta(x_b)}{\tilde{N}_{\lambda}\Big[1-\big\{\frac{\cos(\alpha^2)}{\cosh(\alpha^2)}\big\}^2\Big]}\bigg],\nonumber\\\langle\psi^\pm|\rho^\lambda_{\rm{AB}}|\psi^{\pm}\rangle&=&\frac{1-i^\lambda}{2}\bigg[\frac{1}{2}\pm\frac{\prod\limits_{\delta=0,2}F_\delta(x_a)F_{\delta}(x_b)}{\tilde{N}_{\lambda}\Big[1-\big\{\frac{\cos(\alpha^2)}{\cosh(\alpha^2)}\big\}^2\Big]}\bigg]\nonumber,
\end{eqnarray}
where \begin{eqnarray}
\tilde N_{\lambda=0} &=& \bigg[\frac{F_0(x_a)F_0(x_b)}{1+\frac{\cos(\alpha^2)}{\cosh(\alpha^2)}}\bigg]^2+\bigg[\frac{F_2(x_a)F_2(x_b)}{1-\frac{\cos(\alpha^2)}{\cosh(\alpha^2)}}\bigg]^2,\\
\tilde N_{\lambda=2} &=& \frac{1}{{1-\big\{\frac{\cos(\alpha^2)}{\cosh(\alpha^2)}}\big\}^2}\big[F_0(x_a)^2F_2(x_b)^2\nonumber\\&&+F_2(x_a)^2F_0(x_b)^2\big],
\end{eqnarray}
and $F_\lambda(x) = e^{-\alpha^2}\cosh(2\alpha x)+i^\lambda\cos(2\alpha x)$. 

Fig.\ \ref{fig_4} shows the probability of success and the overlap to the Bell-states $|\phi^\pm\rangle, |\psi^\pm\rangle$ for this implementation of our protocol. We choose $\alpha=1$, and plot the results for the two possible joint-photon-number-modulo-4 measurement outcomes: $\lambda=0,2$, in absence of imperfections and photon loss. For a majority of outcomes in the $(x_a,x_b)$-space, we get one of the four aforementioned Bell-states. The existence of fringes in the plots is due to the measurement of the X-quadratures of both arnie and bert, each of which are in superpositions of $|C_\alpha^+\rangle$ and $|C_{i\alpha}^+\rangle$. The size of the fringes decreases with increasing $\alpha$. The overlaps to the odd (even) Bell-states for the case $\lambda=0(2)$ are identically equal to zero.

So far, we have described our protocol in absence of measurement imperfections and propagation losses. In what follows, we incorporate measurement inefficiencies and propagating losses in our computational model and investigate the resilience of the two different implementations of our protocol to these imperfections. 
\section{Finite Quantum Efficiency and Non-zero Photon Loss}
\label{FinEff}
The dominant source of imperfections in current circuit-QED systems that affect our protocol is undesired photon loss. These losses occur due to photon attenuation on the transmission lines and other lossy devices like circulators and isolators which are necessary for an actual experimental implementation. These lead to decoherence of the entangled states of Alice (Bob) and arnie (bert) as arnie and bert propagate from the stationary  qubits to the  joint-photon-number-modulo-2/4 measurement apparatus and from there on to the homodyne detectors. For simplicity, we assume the losses to be equal for both arnie and bert (the case of unequal losses can be calculated easily using the method described below). Thus, for each of arnie and bert, we define two efficiency parameters $\eta_1$ and $\eta_2$. Here, $\eta_1$ models the losses before the joint-photon-number-modulo-2/4 measurement apparatus and $\eta_2$ models the losses thereafter and before homodyne detection setup. These losses are modeled as photons lost by each of arnie and bert as they pass through beam-splitters with transmission probabilities $\eta_1$ and $\eta_2$ in otherwise perfect transmission lines \cite{Leonhardt_Paul_1995}. Finite qubit lifetimes, with current circuit-QED system parameters, are much less dominant source of imperfection compared to photon loss and are thus neglected in the subsequent analysis.

First, we qualitatively describe the effect of photon loss for the two implementations. Consider the case when the ancilla qubits are encoded in even/odd Schr\"odinger cat states. Loss of a photon is a bit-flip error on the ancilla qubit since $\mathbf{a}|C_\alpha^\pm\rangle\simeq|C_\alpha^\mp\rangle$, where $\mathbf{a}$ is the annihilation operator of the propagating temporal mode. This bit-flip error occurs randomly as the entangled qubit-photon states of Alice (Bob) and arnie (bert) propagate from the stationary qubits to the joint-photon-number-modulo-2 measurement apparatus and from thereon to the homodyne detectors. This results in decoherence of the entangled states of Alice (Bob) and arnie (bert). Therefore, the probability of generating a high fidelity Bell-state of Alice and Bob diminishes drastically (see Sec. \ref{mod 2_imperf}). 

Now, consider the case when the ancillas are encoded in the mod 4 cat states. To lowest order in photon loss, either arnie or bert can lose a photon. On losing a photon, the state of arnie or bert goes from  
$|C_\alpha^{0(2){\rm{mod}}4}\rangle$ to the state $|C_\alpha^{3(1){\rm{mod}}4}\rangle$ (see Eq. \eqref{four_cat_defn_1} and \cite{Mirrahimi_Devoret_2014} for the definition of $|C_\alpha^{1,3{\rm{mod}}4}\rangle$).  Therefore, when either arnie or bert loses a photon, the joint-photon-number-modulo-4 measurement now yields the values $\lambda=1$ or 3, unlike the perfect case outcomes $\lambda=0$ or 2 [see Eqs.\ \eqref{lambda=0_eqn}, \eqref{lambda=2_eqn} and Sec. \ref{mod 4_imperf}]. Thus, measurement of the joint-photon-number-modulo-4 allows us to keep track of loss of a photon in arnie or bert. This tracking of a single photon loss error is equivalent to correcting this error since it allows the knowledge of exact state of the four-qubits after the error has happened. As will be shown below, this enables generation of  high fidelity entangled states of Alice and Bob with higher success rates than the mod 2 implementation. 

In the next order in photon loss, either both arnie and bert lose one photon each or arnie loses two photons, or bert loses two photons. Consider the case in the mod 4 encoding when arnie and bert each lose a photon. Now, the measurement outcome $\lambda$ can be 0 or 2 as in the perfect case and just a measurement of $\lambda$ does not reveal if arnie and bert have indeed lost a photon each. However, these events of loss of one photon each in the ancillas can be tracked by individual photon-number-modulo-2 measurements of the ancillas. In this way, we can suppress the loss of coherence due to loss of a photon in both arnie and bert. Note that the other second order or higher order losses cannot be suppressed by this encoding (more on this in Sec. \ref{concl}). In next two subsections, we describe the effect of photon loss on the the two implementations of our protocol. This is followed by a comparison of the two. Lastly, we describe the protocol with added individual photon-number-modulo-2 measurements of arnie and bert. 

\subsection{Implementation using Schr\"odinger cat states}
\label{mod 2_imperf}
Consider the case when the ancilla qubits are encoded in Schr\"odinger cat states (see Sec. \ref{mod 2_perf}). First, we describe the calculation of the state of Alice and arnie after the propagation of arnie through the transmission line in presence of imperfections. The state of Bob and bert can be computed in an analogous manner. 
Following local entanglement generation between Alice and arnie, their states can be written as:
\begin{eqnarray}
\label{Aa_init_eq}
|\Psi_{\rm{Aa}}\rangle &=& \frac{1}{\sqrt{2}}\big(|g,C_\alpha^+\rangle + |e,C_\alpha^-\rangle\big)\nonumber\\&=&\frac{1}{\sqrt{2}}\sum_{j,\mu=0}^1{\cal N}_j(-1)^{j\mu}|j,(-1)^\mu\alpha\rangle,
\end{eqnarray}
where ${\cal N}_j = 1/\sqrt{2(1+(-1)^j e^{-2|\alpha|^2})}\Rightarrow{\cal N}_{0(1)}={\cal N}_{+(-)}$ and $|j=0(1)\rangle = |g(e)\rangle$.  To compute the decoherence due to propagation losses, without loss of generality, one can introduce an auxiliary propagating mode a$'$, initialized in vacuum, and pass the joint-states of Alice, arnie and a$'$ through a beam-splitter with transmission probability $\eta_1$. Subsequent tracing out of the auxiliary mode yields the reduced density matrix for Alice and arnie after the entangled states of Alice and arnie propagated along the transmission line and  arrived at the joint-photon-number-modulo-2 apparatus. 

Therefore, just prior to making the joint-photon-number-modulo-2 measurement, we can write the total state of the four-modes to be (cf. Appendix. \ref{appAmod2_1} for details of the calculation) : 
\begin{eqnarray}
\label{rhoABab_eqn}
\rho_{\rm{ABab}} &=& \frac{1}{64}\sum_{\rm{Aa, Bb}}\frac{{\cal N}_j{\cal N}_{j'}{\cal N}_l{\cal N}_{l'}}{\bar{\cal N}_k\bar{\cal N}_{k'}\bar{\cal N}_m\bar{\cal N}_{m'}}(-1)^{\bm{\mu}\cdot(\mathbf{j}+\mathbf{k}) +\bm{\nu}\cdot(\mathbf{l}+\mathbf{m})}\nonumber\\&&e^{-\epsilon^2\{2-(-1)^{\mu+\mu'}-(-1)^{\nu+\nu'}\}}\big(|j,l\rangle\langle j',l'|\big)\nonumber\\&&\big|C_\alpha^{(-)^k}, C_\alpha^{(-)^m}\big\rangle\big\langle C_\alpha^{(-)^{k'}}, C_\alpha^{(-)^{m'}}\big|,
\end{eqnarray}
where $\sum\limits_{\rm{Aa}} = \sum\limits_{j,j'=0}^1\sum\limits_{k,k'=0}^1\sum\limits_{\mu,\mu'=0}^1$, $\sum\limits_{\rm{Bb}} =\sum\limits_{l,l'=0}^1\sum\limits_{m,m'=0}^1\sum\limits_{\nu,\nu'=0}^1, \bm{\mu}=\{\mu,\mu'\}, \bm{\nu}=\{\nu,\nu'\}, \mathbf{j}=\{j,j'\},\mathbf{k}=\{k,k'\}, \mathbf{l}=\{l,l'\}, \mathbf{m}=\{m,m'\}$. Furthermore, we have defined  $\bar{\cal N}_j = 1/\sqrt{2(1+(-1)^j e^{-2|\bar\alpha|^2})}$ where  $\bar\alpha=\sqrt{\eta_1}\alpha, \epsilon = \sqrt{1-\eta_1}\alpha$. 

Note that Eq.\ \eqref{rhoABab_eqn} is expressed in the eigenbasis of the joint-photon-number-modulo-2 measurement. An outcome of $p=1(-1)$ results in the state of Alice, Bob, arnie and bert to be in $\rho_{\rm{ABab}}^{p}=\bar\rho_{\rm{ABab}}^{p}/{\rm{Tr}}[\bar\rho_{\rm{ABab}}^{p}]$, where the post-measurement un-normalized density matrix $\bar\rho_{\rm{ABab}}^{p}$ is given by: 
\begin{eqnarray}
\bar\rho_{\rm{ABab}}^{p} &=& \frac{1}{64}\sum_{\rm{Aa, Bb}}'\frac{{\cal N}_j{\cal N}_{j'}{\cal N}_l{\cal N}_{l'}}{\bar{\cal N}_k\bar{\cal N}_{k'}\bar{\cal N}_m\bar{\cal N}_{m'}}(-1)^{\bm{\mu}\cdot(\mathbf{j}+\mathbf{k}) +\bm{\nu}\cdot(\mathbf{l}+\mathbf{m})}\nonumber\\&&e^{-\epsilon^2\{2-(-1)^{\mu+\mu'}-(-1)^{\nu+\nu'}\}}\big(|j,l\rangle\langle j',l'|\big)\nonumber\\&&\big|C_\alpha^{(-)^k}, C_\alpha^{(-)^m}\big\rangle\big\langle C_\alpha^{(-)^{k'}}, C_\alpha^{(-)^{m'}}\big|.
\end{eqnarray}
Here the prime in $\sum\limits_{\rm{Aa, Bb}}'$ indicates that the summation has to be performed while keeping $k+m=0(1)\ {\rm{mod}}\ 2, k'+m'=0(1)\ {\rm{mod}}\ 2$ for an outcome of $p=1(-1)$. 

Next, we treat the photon-losses after the joint-photon-number-modulo-2 measurement setup and before the homodyne detectors. We model the measurement operator for the imperfect homodyne detection with efficiency $\eta_2$ as a superposition of projectors $|x_a,x_b\rangle\langle x_a, x_b|$ with a Gaussian envelope \cite{Leonhardt_Paul_1995, Steck_2015}:
 \begin{eqnarray}
 \label{omegaq}
\Omega_{Q} &=& \frac{1}{\sigma\sqrt{\pi\eta_2}}\int_{-\infty}^\infty dx_a\int_{-\infty}^\infty dx_b\ e^{-\frac{1}{2\sigma^2}(\frac{q_a}{\sqrt{\eta_2}}-x_a)^2}\nonumber\\&&e^{-\frac{1}{2\sigma^2}(\frac{q_b}{\sqrt{\eta_2}}-x_b)^2}|x_a,x_b\rangle\langle x_a,x_b|,\nonumber\\\label{ineff_hom}\sigma^2 &=& \frac{1-\eta_2}{2\eta_2}.
\end{eqnarray}
Here $q_{a(b)}$ denote the imperfect measurement outcome which, in principle, can arise out of the possible perfect measurement outcomes $x_{a(b)}$ with a Gaussian probability distribution shown above. For $p=\pm1$, the system density matrix due to this measurement evolves as: $\rho^p_{\rm{ABab}} \rightarrow \Omega_{Q}\rho^p_{\rm{ABab}} \Omega_Q^\dagger/{\rm Tr}\big[\Omega_Q\rho^p_{\rm{ABab}} \Omega_Q^\dagger\big]$, where the product ${\rm{Tr}}[\bar\rho_{\rm{ABab}}^{p}]\times{\rm Tr}\big[\Omega_Q\rho^p_{\rm{ABab}} \Omega_Q^\dagger\big]$ is the probability distribution of outcomes $\bar{P}^p(q_a, q_b)$. These can be evaluated analytically following the method outlined in Appendix. \ref{appAmod2}, along with the overlap to the Bell-states $|\phi^\pm\rangle, |\psi^\pm\rangle$. The explicit forms of these quantities are not provided for the sake of brevity. 

\begin{figure}
\centering
\includegraphics[width = 0.5\textwidth]{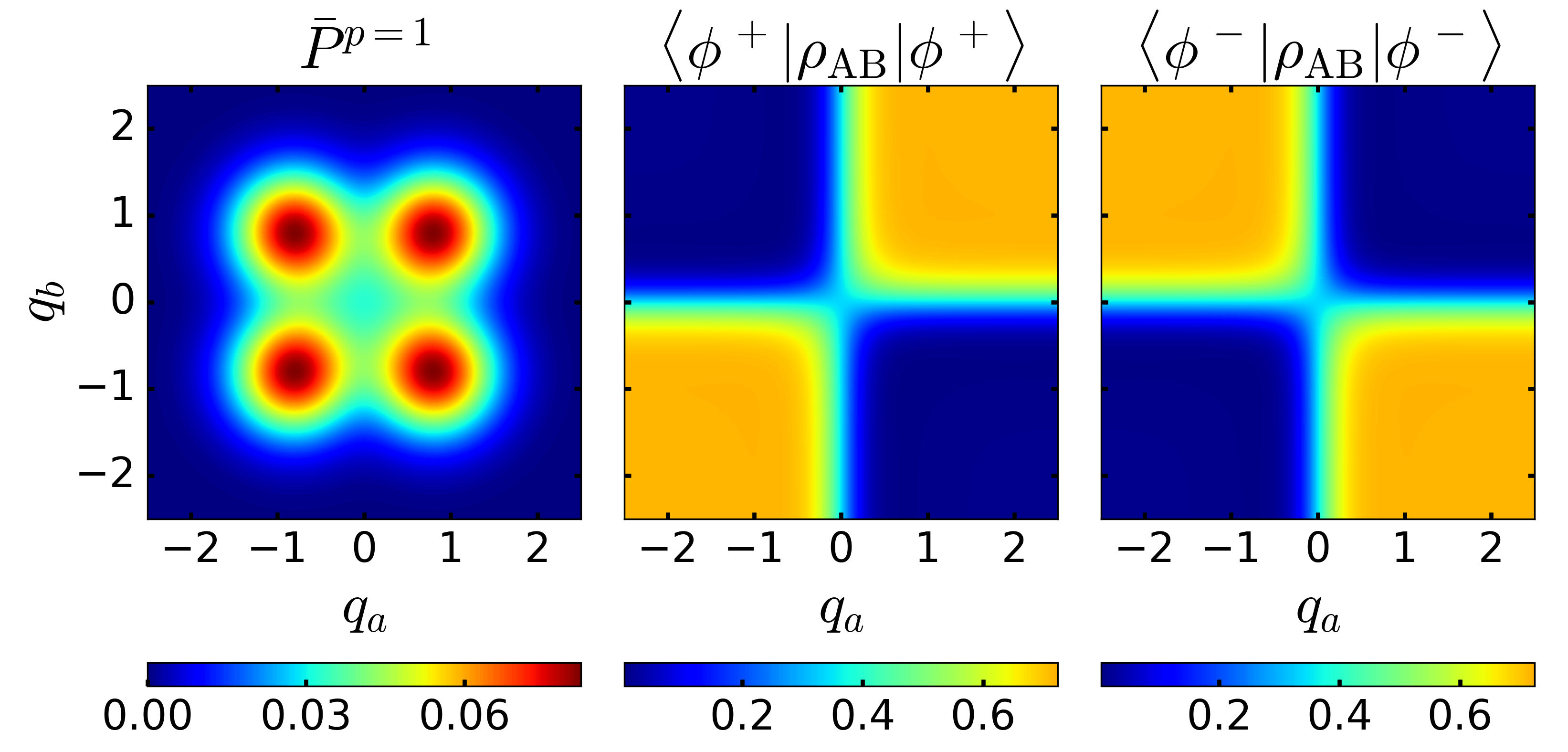}
\caption{ \label{fig_5} (color online) Probability distribution $\bar{P}^p(q_a, q_b)$ of outcomes of the homodyne measurements of arnie and bert and resulting overlaps of Alice and Bob's joint density matrix $\rho_{AB}$ to the Bell-states $|\phi^\pm\rangle$ are shown for the case when the joint-photon-number-modulo-2 measurement yields $p=1$. We choose $\alpha=1$ and $\eta_1 = \eta_2=0.8$. (Left) Probability distribution showing four Gaussian distributions centered at $q_a = \pm\bar{\alpha}, q_b = \pm\bar{\alpha}$. (Center and Right) The corresponding overlap to the Bell-state $|\phi^+\rangle$ is substantial for $(q_a, q_b)$ in the vicinity of $(\bar\alpha,\bar\alpha)$ and $(-\bar\alpha,-\bar\alpha)$, while the overlap to the Bell-state $|\phi^-\rangle$ is substantial for $(q_a, q_b)$ in the vicinity of $(\bar\alpha,-\bar\alpha)$ and $(-\bar\alpha,\bar\alpha)$. We note that the maximum fidelity Bell-state that can be obtained for outcomes with non-negligible occurrence probability is $\sim0.7$ instead of 1.0 that was obtained for perfect efficiency (compare Fig.\ \ref{fig_3}). Outcomes along the lines $q_a=0$ and $q_b=0$ do not yield entangled states for reasons similar to that given for perfect efficiency. Similar results hold for the case $p=-1$. }
\end{figure}

Fig.\ \ref{fig_5} shows the probability distribution of outcomes for the imperfect measurement $\bar{P}^p(q_a, q_b)$ and the overlaps to the Bell-states $|\phi^\pm\rangle$ for the case when the joint-photon-number-modulo-2 measurement yields $p=1$. We choose $\alpha=1$ and the measurement inefficiencies to be: $\eta_1= \eta_2=0.8$. The probability distribution shows four Gaussian distributions centered at $q_a = \pm\bar{\alpha}, q_b = \pm\bar{\alpha}$. The corresponding overlap to the Bell-state $|\phi^+\rangle$ is substantial for $(q_a, q_b)$ in the vicinity of $(\bar\alpha,\bar\alpha)$ and $(-\bar\alpha,-\bar\alpha)$, while the overlap to the Bell-state $|\phi^-\rangle$ is substantial for $(q_a, q_b)$ in the vicinity of $(\bar\alpha,-\bar\alpha)$ and $(-\bar\alpha,\bar\alpha)$. Note that in presence of imperfections, the maximum fidelity for outcomes with non-negligible occurrence probability is $\sim0.7$ instead of 1.0 computed earlier for the case of perfect efficiency (compare Fig.\ \ref{fig_3}).  For outcomes along the lines $q_a=0$ and $q_b=0$, one does not generate entangled states for reasons similar to the case of perfect efficiency.
Similar set of results hold for $p=-1$. 
  
\subsection{Implementation using mod 4 cat states}
\label{mod 4_imperf}
In this section, we consider the case when the ancilla qubits are encoded in the mod 4 cat states discussed in Sec. \ref{mod 4_perf}. The computation for this case is, in principle, similar to that outlined in the previous subsection. We begin by considering the entangled qubit-photon state of Alice and arnie, which, in this case, is given by: 
\begin{eqnarray}
\label{Aa_init_eq_mod4}
|\Psi_{\rm{Aa}}\rangle &=& \frac{1}{\sqrt{2}}\big(|g,C_\alpha^{0{\rm{mod}}4}\rangle + |e,C_\alpha^{2{\rm{mod}}4}\rangle\big)\nonumber\\&=&\frac{1}{\sqrt{2}}\sum_{j,\mu,\nu=0}^1\tilde{\cal N}_{2j}{\cal N}_{2j}(-1)^{j\nu}|j,(-1)^\mu i^\nu\alpha\rangle,
\end{eqnarray}
where ${\cal N}_{2j}$ and the states $|j=0,1\rangle$ for Alice are defined as before. The definition of $\tilde{\cal N}_{2j}$ is obtained by setting $\lambda =2j$ in the following: $\tilde{\cal N}_\lambda = \big[2+2(-i)^\lambda\{e^{i|\alpha|^2}+(-1)^\lambda e^{-i|\alpha|^2}\}/
\{e^{|\alpha|^2}+(-1)^\lambda e^{-|\alpha|^2}\}\big]^{-\frac{1}{2}}$, $\lambda = \{0,1,2,3\}$. To compute the state of Alice and arnie after their entangled qubit-photon state encounter propagation losses, we follow a similar approach as in the previous section: introducing an auxiliary mode a$'$, looking at the resultant state after passage through a beam-splitter with transmission probability $\eta_1$ and subsequently, tracing out the mode a$'$. Similar set of calculations can be done for Bob and bert. Performing this computation results in (cf. Appendix \ref{appAmod4_1} for details): 
\begin{eqnarray}
\label{ABab_mod_4_imperf}
\rho_{\rm{ABab}} &=& \frac{1}{2^{10}}\sum\limits_{\rm{Aa,Bb}}\frac{\prod\limits_l\tilde{\cal N}_{2l}{\cal N}_{2l}}{\prod\limits_\zeta\bar{\tilde{\cal N}}_{\zeta}\bar{\cal N}_{\zeta}}(-1)^{\bm{\mu}\cdot\bm{\gamma}+\bm{\nu}\cdot\bm{j}+\bm{\phi}\cdot\bm{\delta}+\bm{\psi}\cdot\bm{k}}i^{\nu\gamma-\nu'\gamma'}\nonumber\\&&i^{\psi\delta-\psi'\delta'}e^{-\epsilon^2\{2-(-1)^{\mu+\mu'}i^{\nu-\nu'}-(-1)^{\phi+\phi'}i^{\psi-\psi'}\}}\big(|j,k\rangle\nonumber\\&&\langle j'k'|\big)\Big|C_{\bar\alpha}^{\gamma{\rm{mod}}4},C_{\bar\alpha}^{\delta{\rm{mod}}4}\Big\rangle\Big\langle C_{\bar\alpha}^{\gamma'{\rm{mod}}4},C_{\bar\alpha}^{\delta'{\rm{mod}}4}\Big|,
\end{eqnarray}
where $\prod\limits_l=\prod\limits_{l\in\{j,j',k,k'\}}, \prod\limits_\zeta = \prod\limits_{\zeta\in\{\delta,\delta',\gamma,\gamma'\}}, \sum\limits_{\rm{Aa}} = \sum\limits_{j,j'=0}^1 \sum\limits_{\mu,\mu'=0}^1 \sum\limits_{\nu,\nu'=0}^1\sum\limits_{\gamma,\gamma'=0}^3, \sum\limits_{\rm{Bb}} = \sum\limits_{k,k'=0}^1 \sum\limits_{\phi,\phi'=0}^1 \sum\limits_{\psi,\psi'=0}^1\sum\limits_{\delta,\delta'=0}^3$ and $\bm{j} = \{j,j'\}, \bm{\mu} = \{\mu,\mu'\}, \bm{\nu}=\{\nu,\nu'\}$ and $\bm{\gamma} = \{\gamma, \gamma'\}, \bm{k} = \{k,k'\}, \bm{\phi} = \{\phi,\phi'\}, \bm{\psi}=\{\psi,\psi'\}$ and $\bm{\delta} = \{\delta, \delta'\}$. The definitions of $\bar{\tilde{\cal N}}_\gamma,\bar{\cal N}_\gamma$ can obtained from the definitions of ${\tilde{\cal N}}_\gamma,{\cal N}_\gamma$ (cf. Secs. \ref{mod 2_imperf}, \ref{mod 4_imperf}) by making the substitution $\alpha\rightarrow\bar\alpha$ and $\bar\alpha, \epsilon$ have been defined in the previous subsection. 

Noting that Eq.\ \eqref{ABab_mod_4_imperf} expresses the density matrix in the eigenbasis of the joint-photon-number-modulo-4 measurement of arnie and bert, an outcome $\lambda\in\{0,1,2,3\}$ projects the state of the four modes to $\rho_{\rm{ABab}}^\lambda = \bar\rho_{\rm{ABab}}^\lambda/{\rm{Tr}}\big[\bar\rho_{\rm{ABab}}^\lambda\big]$, where $\bar\rho_{\rm{ABab}}^\lambda$ is the un-normalized density matrix obtained from $\rho_{\rm{ABab}}$ by restricting the summation of $\gamma,\gamma', \delta, \delta'$ to be such that: $\gamma+\delta=\lambda\ {\rm{mod}}\ 4, \gamma'+\delta'=\lambda\ {\rm{mod}}\ 4$. The inefficiencies in the final homodyne detection of arnie and bert can be done similarly to the method described in Sec. \ref{mod 2_imperf}, using Eq.\ \eqref{ineff_hom}. The probability of outcomes and the overlap to the Bell-states $|\phi^\pm\rangle,|\psi^\pm\rangle$ can be evaluated analytically, whose explicit forms are not shown here for brevity. 

\begin{figure}
\centering
\includegraphics[width = 0.5\textwidth]{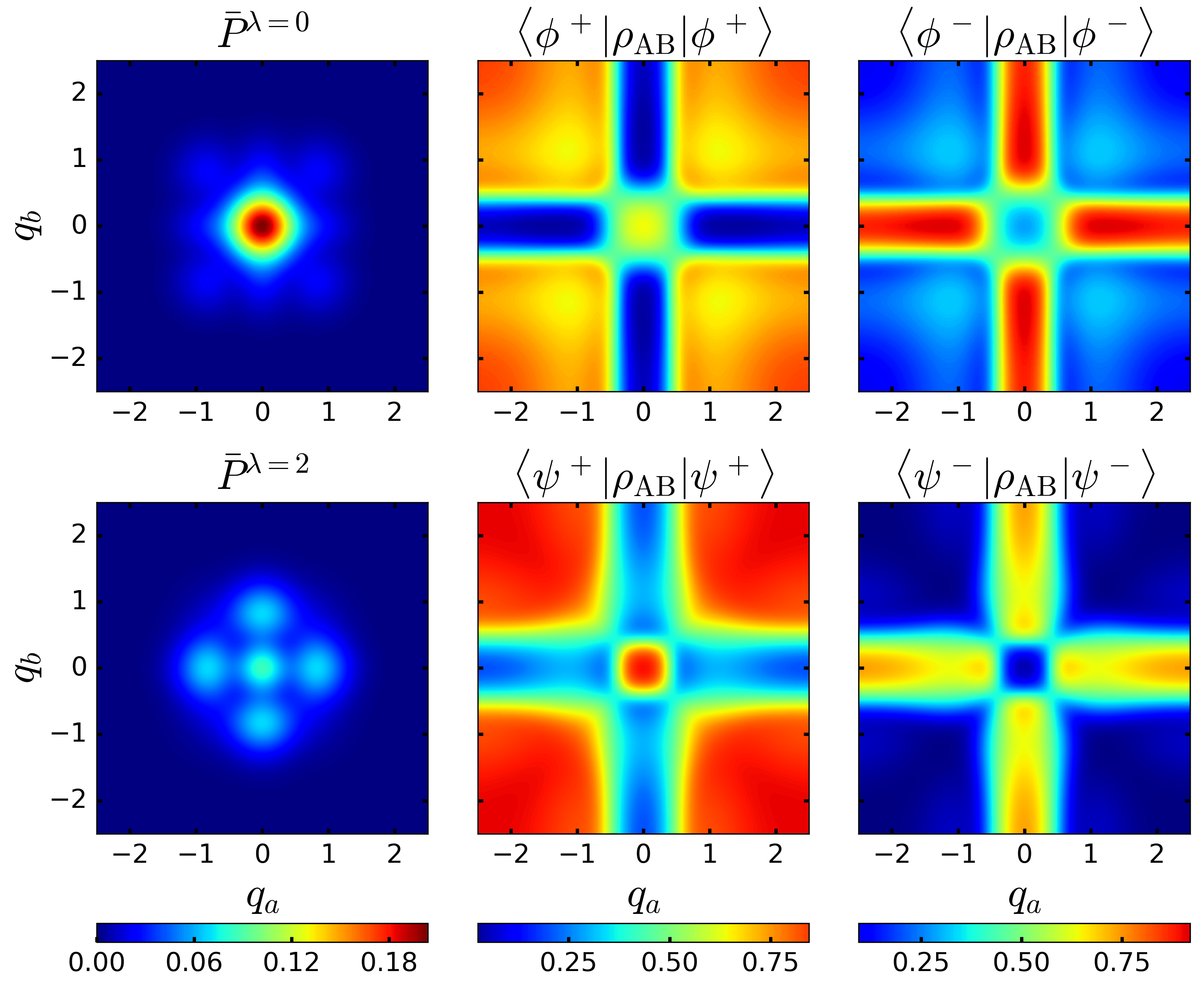}
\caption{ \label{fig_6} (color online) Probability distribution $\bar{P}^\lambda(q_a, q_b)$ of outcomes of the homodyne measurements of arnie and bert and resulting overlaps of Alice and Bob's joint density matrix $\rho_{\rm{AB}}$ with the Bell-states $|\phi^\pm\rangle = (|g,g\rangle \pm|e,e\rangle)/\sqrt{2}, |\psi^\pm\rangle = (|g,e\rangle \pm|e,g\rangle)/\sqrt{2}$ are shown. We choose $\alpha=1.0$ and $\eta_1 = \eta_2=0.8$ and show the cases $\lambda=0,2$ (see Fig.\ \ref{fig_6_a} in Appendix. \ref{lambda_1_3} for $\lambda=1,3$). The top (bottom) left panel shows the probability of outcomes for the joint-photon-number-modulo-4 outcome to be $\lambda=0(2)$. Corresponding overlaps to the Bell-states $|\phi^\pm\rangle(|\psi^\pm\rangle)$ are plotted in the top (bottom) center and top (bottom) right panels. Note that the maximum fidelity obtained for outcomes with non-negligible occurrence probability is lowered compared to the perfect case (compare Fig.\ \ref{fig_4}). 
The overlap to the odd (even) Bell-states for $\lambda=0(2)$ are not shown for brevity. }
\end{figure}
Fig.\ \ref{fig_6} shows the probability of outcomes and the overlap to the Bell-states $|\phi^\pm\rangle(|\psi^\pm\rangle)$ when the joint-photon-number-modulo-4 measurement outcome $\lambda = 0(2)$. For brevity, the results for the outcomes $\lambda=1,3$, (the cases where the loss of a single photon in either arnie or bert was tracked) are shown in Fig.\ \ref{fig_6_a} in Appendix. \ref{lambda_1_3}. We have chosen $\alpha=1.0$ and $\eta_1=\eta_2=0.8$. The top (bottom) left panel shows the probability of outcomes for $\lambda = 0(2)$, while the top (bottom) center and right panels show the corresponding overlaps to the Bell-states $|\phi^\pm\rangle(|\psi^\pm\rangle)$. We see that including inefficiencies lowers the maximum fidelity obtained for outcomes with non-negligible occurrence probability compared to perfect case (compare Fig.\ \ref{fig_4}). 

\subsection{Comparison of the mod 2 and mod 4 implementations in presence and absence of imperfections}
\label{mod 2_mod 4_imperf_comp}
In the previous subsections, we described the probability of outcomes for the two different implementations of our protocol. In this section, we discuss the the total and optimized success-rates of generating entangled states for the two implementations, comparing the cases of perfect and imperfect quantum efficiencies. 

First, consider the mod 2 implementation of our protocol for perfect and imperfect quantum efficiencies. The probability of success and the overlaps to the Bell-states are given in Fig.\ \ref{fig_3} (Fig.\ \ref{fig_5}) for the perfect (imperfect) case. 
The total success-rate for generation of entangled states can be computed for different cut-off fidelities by integrating the appropriate region of $(x_a,x_b)$ or $(q_a,q_b)$ space of outcomes. In the perfect (imperfect) case, the majority of the outcomes, occurring around $\pm\alpha(\pm\bar\alpha)$, give rise to entangled states, while the events along the lines $x_a(q_a)=0$ and $x_b(q_b)=0$ do not. Thus, in order to have a high total success-rate of generating entangled states, the number of outcomes along the lines $x_a(q_a)=0$ and $x_b(q_b)=0$ should be minimized. This can be done by increasing the size of $\alpha$ because the probability of obtaining an outcome along these lines goes down exponentially with $\alpha^2(\bar\alpha^2)$. While in the perfect case $\alpha$ can be made arbitrarily large giving rise to deterministic generation of entangled states, in the imperfect case, too large a value of $\alpha$ lowers the success-rate. This is because large values of $\alpha$ are more susceptible to photon-losses. 

Next, consider the mod 4 implementation of our protocol. The relevant probability of outcomes and overlap to the Bell-states are given in Fig.\ \ref{fig_4} (Fig.\ \ref{fig_6}) for the perfect (imperfect) case. Similar considerations, as in the mod 2 implementation, lead to the conclusion that in absence of imperfections, increasing $\alpha$, in general, increases the success-rate for the different cut-off fidelities. Note that the increase, however, is not monotonic (see below). However, in presence of imperfections, arbitrarily increasing $\alpha$ does not increase the success-rate of generating high-fidelity entangled states. This happens again because large values of $\alpha$ are more susceptible to photon-losses. However, for large values of $\alpha$, when photon loss dominates, tracking the loss of photons and thereby correcting the errors enables higher success rate of generating better entangled state compared to the mod 2 implementation. 

\begin{figure}[!h]
\centering
\includegraphics[width = 0.5\textwidth]{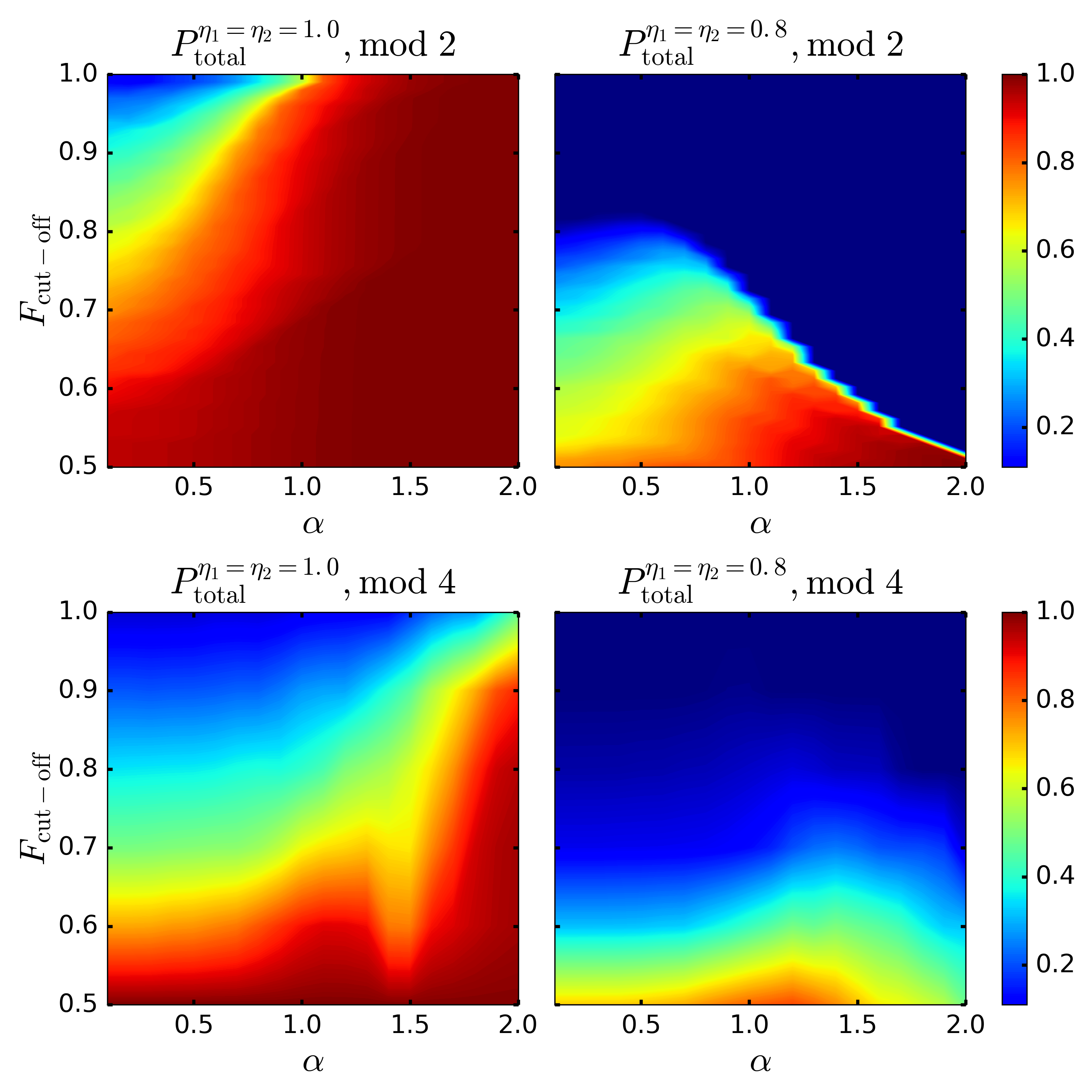}
\caption{ \label{fig_7} (color online) Total success probability ($P_{\rm{total}}$) for different cut-off fidelities and different choices of $\alpha$ is plotted for the case of perfect quantum efficiency (left panels, indicated by $\eta_1 = \eta_2 = 1$) and the imperfect case (right panels, where we have chosen $\eta_1 = \eta_2 = 0.8$). The top (bottom) panels correspond to the mod 2 (mod 4) implementation. (Top left) For $\alpha\ll1$, the probability of generation of entangled states with overlap $>0.9$ is around $0.5$. Increasing $\alpha$ to $\gg1$ generates perfect entangled states with near-unit probability. (Top right) In presence of imperfections, for $\alpha\ll1$, we generate entangled states with overlap $>0.7$ to Bell-states with probability in excess of $0.3$. However, increasing $\alpha$ does not lead to a higher success-rate for generating better entangled states. This is because larger values of $\alpha$ are more susceptible to photon loss. Depending on the desired cut-off fidelity and the efficiency of an experimental setup, there is an optimal choice of $\alpha$ that leads to the maximal success-rate. For instance, in the case shown, for a desired cut-off fidelity of $0.75$, the optimal choice is $\alpha\simeq0.7$. (Bottom left) The total probability of generating entangled states with overlap $>0.9$ is $\sim0.3$ and increasing $\alpha$ increases the success-rate to near-unity. The increase is non-monotonic because the size of the fringes in the overlap (Figs. \ref{fig_4}, \ref{fig_6}) depend on the value of $\alpha$. (Bottom right) In presence of imperfections, increasing $\alpha$ to $\gg1$ no longer increases the success-rate. As in the mod 2 case, there is an optimal choice for $\alpha$: e.g. for a cut-off fidelity of $0.75$ for this choice of inefficiency, $\alpha\sim1.5$. Further, for relatively large values of $\alpha\geq1.5$, it is more advantageous to use the mod 4 implementation over the mod 2 implementation. This is because the mod 4 protocol corrects for the decoherence of the entangled qubit-photon states due to photon-loss to first order. }
\end{figure}

Fig.\ \ref{fig_7} shows the total probability of generation of entangled states as a function of different cut-off fidelities and different choices of the parameter $\alpha$ for the perfect case (left panels,  indicated by $\eta_1=\eta_2=1$) and the imperfect case (right panels, for which, we have chosen $\eta_1 = \eta_2 = 0.8$). The top (bottom) panels correspond to the mod 2 (mod 4)  implementation. For the mod 2 implementation, for the perfect case, for $\alpha\ll1$, the probability of generation of entangled states with an overlap $>0.9$ to a Bell-state is $\sim0.5$ while for $\alpha>1$ for which we generate entangled states with unit-probability. On the other hand, in the imperfect case, for the choice of efficiency parameters $\eta_1 = \eta_2 = 0.8$, small values of $\alpha$ ($\alpha\ll1$) give rise to entangled states with overlaps to Bell-states $>0.7$ with a success-rate in excess of $0.3$. However, unlike the perfect case, larger values of $\alpha$ do not help getting better success-rate for similar or better entangled states because of photon loss. Thus, for different cut-off fidelities and measurement efficiencies, there is an optimal choice of $\alpha$, e.g. in the case shown, for a cut-off fidelity of $0.75$, the optimal choice for $\alpha$ is $\simeq0.7$. For the mod 4 implementation, the success-rates are $\sim0.3$ for generating entangled states with overlap $>0.9$ in absence of imperfections for $\alpha\ll 1$. Increasing $\alpha$ increases the success-rate to near-unity. Note that the increase is non-monotonic, unlike the case of mod 2 implementation. This is because the size of the fringes, which are regions of unentangled states, present in the overlap to the Bell-states (see Figs. \ref{fig_4}, \ref{fig_6}) depends on the value of $\alpha$. In presence of imperfections, increasing $\alpha$ does not necessarily increase the success-rate because of decoherence due to photon loss. Just as in the mod 2 implementation, there is an optimal choice for $\alpha$: e.g. for a cut-off fidelity of $0.75$ for this choice of inefficiency, $\alpha\sim1.5$. Note that for relatively larger values of $\alpha>1.5$, it is more advantageous to use the mod 4 implementation over the mod 2 implementation since it corrects for the decoherence of the propagating qubit-photon states due to photon-loss to first order. 

Next, we optimize the success-rate with respect to the parameter $\alpha$ for the two implementations. This optimization is done numerically for the different values of the efficiency parameters $\eta_1,\eta_2$ and the different cut-off fidelities. This is shown below in Fig.\ \ref{fig_8}. We take $\eta_1=\eta_2$ for simplicity. 
For $\eta_1=\eta_2 =0.8$, one is able to generate entangled states with overlap to Bell-states $\sim0.75$ with a near-unity success-rate by both mod 2 and mod 4 implementation. Although the probability of generation of higher fidelity Bell-states decreases for both the implementations, the rate of decrease is different for the two. Enclosed by the white curves in the right panel is the region where the mod 4 implementation has a higher success rate than the mod 2 implementation. For instance, for $\eta_1=\eta_2=0.9$, the probability of generating a Bell-state with overlap of $0.95$ or greater is less than $10^{-10}$ for the mod 2 implementation (white rectangle in left panel). On the other hand, the mod 4 implementation is able to generate these states with a success-rate of $10^{-4}$ (white rectangle in right panel). This is because of its ability to correct for photon loss errors to first order. Note, however, that for low enough efficiency and high enough cut-off fidelity, the error correcting mod 4 protocol ceases to be advantageous. This is because for such low efficiency because higher than first order photon loss become more dominant. 
\begin{figure}
\centering
\includegraphics[width = 0.5\textwidth]{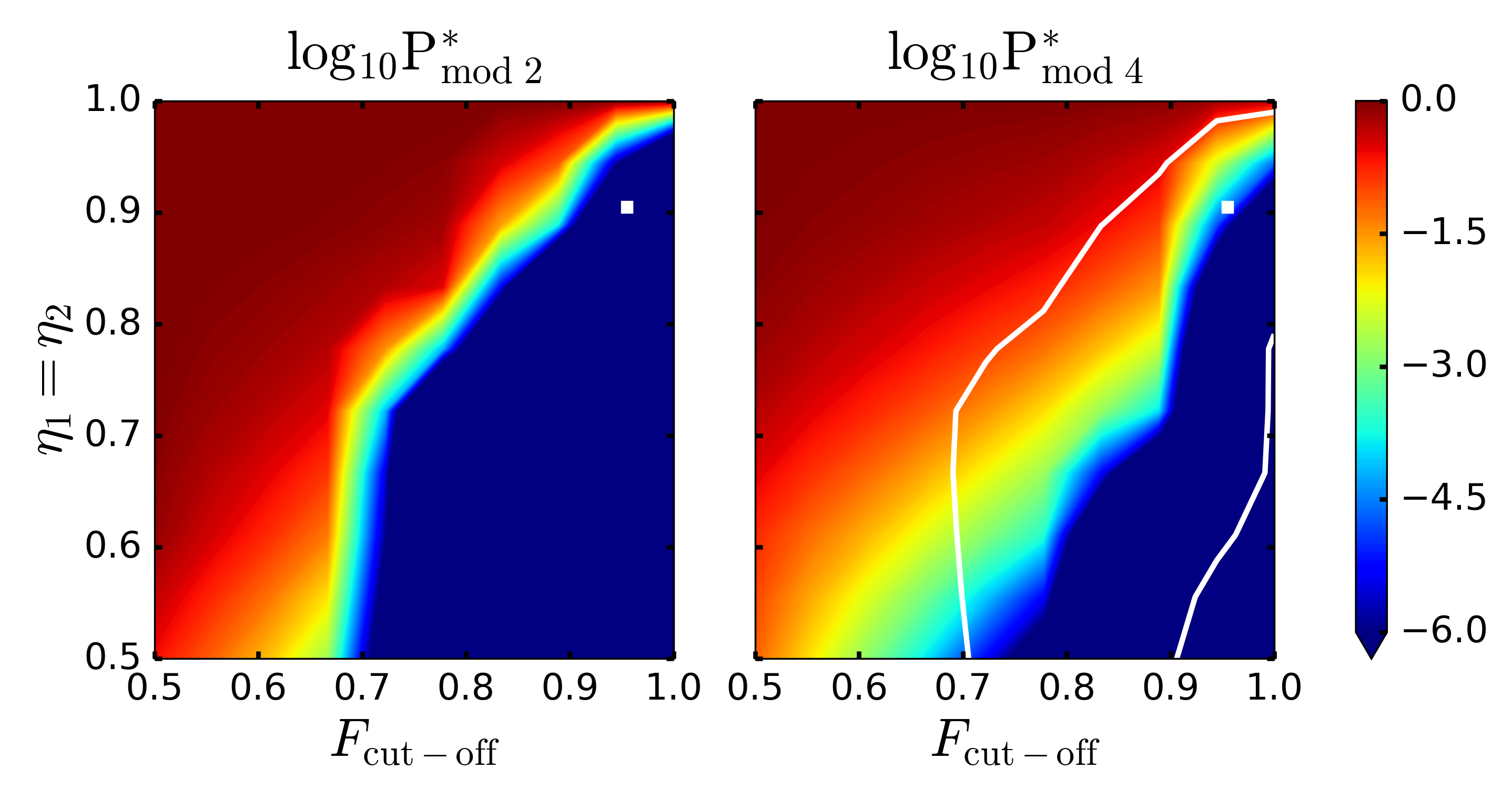}
\caption{ \label{fig_8} (color online) (Top panels) Optimized total success-probability $P^*$ in logarithmic scale for different cut-off fidelities and different choices of $\eta_1,\eta_2$ for the mod 2 and the mod 4 implementations. The optimization is done numerically for the value of the parameter $\alpha$. We choose, for simplicity, $\eta_1=\eta_2$. The left (right) panel corresponds to the joint-photon-number-modulo-2(4) implementation. For both protocols, for an efficiency of $\eta_1 = \eta_2 = 0.8$, entangled states with overlaps to the Bell-states $\sim0.75$ are generated with near-unity success-rate. The probability of generating high-fidelity  Bell-states diminishes rapidly. However, the rate of decrease of success-rate for the mod 2 and mod 4 implementations are different. The white curves in the right panel enclose the region for which the mod 4 implementation has a higher success rate than the mod 2 implementation. For instance, for inefficiency values $\sim 0.9$, the success-rate for the mod 2 implementation is less than $10^{-10}$ for generating states with overlap to Bell-states of $\sim0.95$ (white rectangle in left panel). However, the mod 4 implementation, due to its ability to correct for photon loss errors to first order, can, in fact, generate states with overlap $\sim0.95$ to Bell-states with a success-rate of $10^{-4}$ (white rectangle in right panel). However, the error correcting mod 4 protocol ceases to be advantageous to generate high fidelity Bell-states for low enough efficiencies and high enough cut-off fidelities. This is because for such low efficiency, higher order photon loss become more dominant.}
\end{figure}

\subsection{Adding individual photon-number-modulo-2 measurements to the mod 4 implementation}
\label{mod 4_p1p2}
In the previous subsections, we have shown that in presence of finite quantum efficiency, it is more advantageous to use the mod 4 implementation of our protocol, because this implementation corrects for decoherence due to loss of a photon in either arnie or bert. In this section, we describe an improvement of the mod 4 implementation. The improvement consists of measurement of individual photon-number-modulo-2 of each of arnie and bert, in addition to the joint-photon-number-modulo-4 measurement. Thus, this improved mod 4 implementation is referred to as the (mod 4) $+ \rm{P}_a + \rm{P}_b$ implementation, where $\rm{P}_{a,b}$ denote the individual photon-number-modulo-2 measurements of arnie, bert. 
As explained in the previous subsection, this improvement suppresses decoherence due to the loss of one photon in both arnie and bert and increases the success-rate of generating high fidelity entangled Bell-states compared to the mod 4 implementation. Note that in absence of imperfections, the measurement of the individual parity of arnie and bert provides no additional information and advantage. 

Incorporating this improvement in an experimental implementation poses no additional challenge compared to the mod 4 implementation as demonstrated in \cite{Wang_Schoelkopf_2016}. Further, the time required to make these additional measurements, with current circuit-QED parameters, is negligible compared to the typical qubit coherence times. This justifies neglecting the qubit decoherence for this part of the analysis. 
The theoretical calculations can be done in an analogous manner to that described in Sec. \ref{mod 4_imperf}. The only difference comes in while computing the resultant state of Alice, Bob, arnie and bert after the individual photon-number-modulo-2 and the joint-photon-number-modulo-4 measurements, given by $\rho_{\rm{ABab}}^\lambda = \bar\rho_{\rm{ABab}}^\lambda/{\rm{Tr}}\big[\bar\rho_{\rm{ABab}}^\lambda\big]$. As before, the un-normalized density matrix $\bar\rho_{\rm{ABab}}^\lambda$ is obtained from $\rho_{\rm{ABab}}$ [given in Eq. \eqref{ABab_mod_4_imperf}] by restricting the summation of $\gamma,\gamma',\delta, \delta'$ to be such that: $\gamma+\delta=\lambda\ {\rm{mod}}\ 4, \gamma'+\delta'=\lambda\ {\rm{mod}}\ 4$. The only difference is that depending on the individual parity of arnie (bert) to be even or odd, the values of $\gamma,\gamma' (\delta, \delta')$ are restricted to $0,2$ or $1,3$. The computation of the homodyne detection can also be done as before. 

\begin{figure}
\centering
\includegraphics[width = 0.5\textwidth]{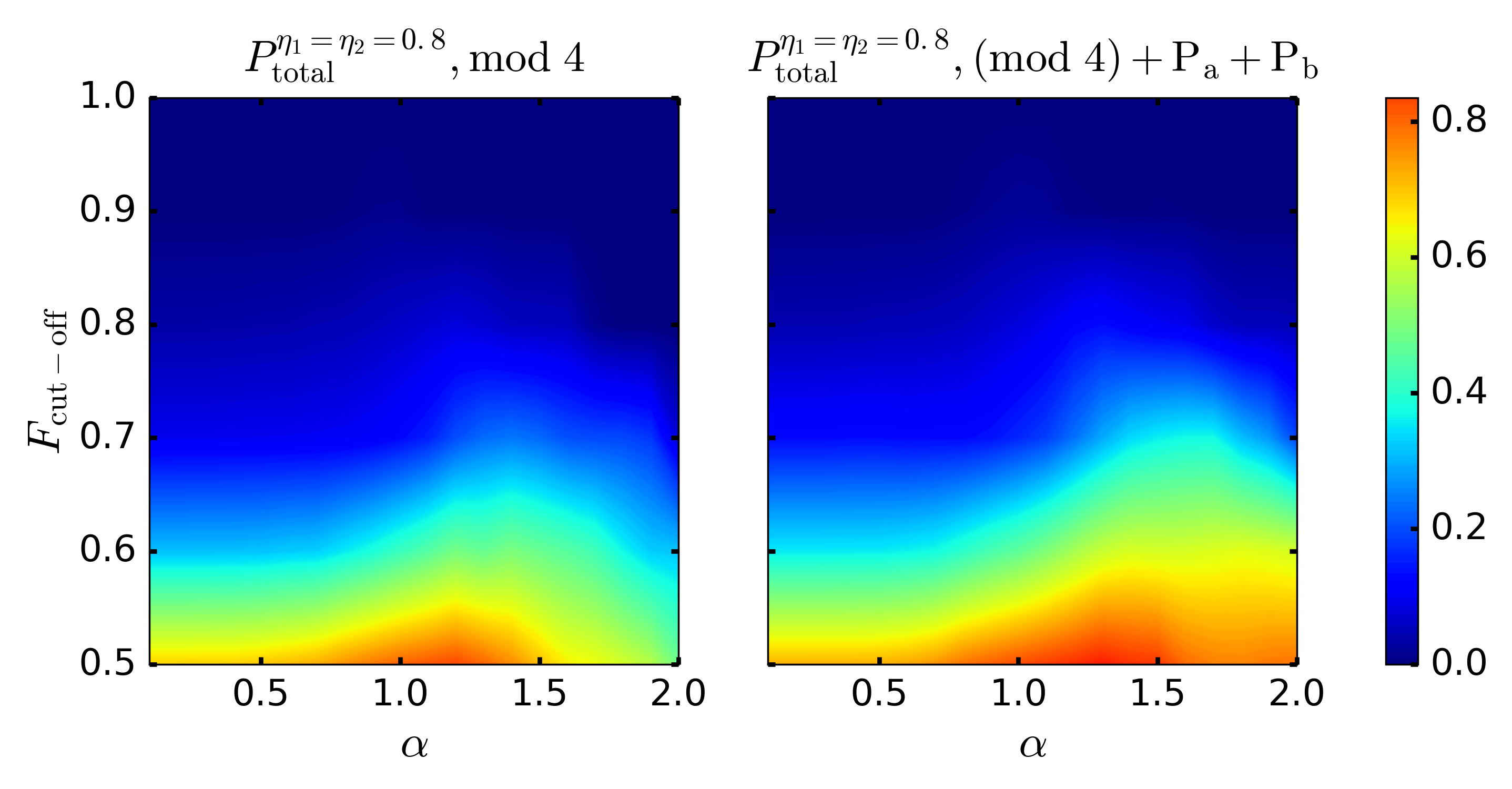}
\caption{ \label{fig_9} (color online) Total success rate for the mod 4 (left panel) and the (mod 4) $+ \rm{P}_a + \rm{P}_b$ (right panel) implementations are shown for the case of finite quantum efficiency, where we choose $\eta_1=\eta_2 = 0.8$. While for $\alpha\ll1$, both the mod 4 and the (mod 4) $+ \rm{P}_a + \rm{P}_b$ implementation perform similarly, for larger values of $\alpha$, the latter performs better than the other. This is because the mod 4 implementation corrects loss of a single photon loss in either arnie or bert, while the  (mod 4) $+ \rm{P}_a + \rm{P}_b$ implementation corrects for the loss of single photons in each of arnie and bert. }
\end{figure}

The results of the computation for the optimized total success-rate is plotted in Fig.\ \ref{fig_9}. We show total success rate for the (mod 4) $+ \rm{P}_a + \rm{P}_b$ implementation in the case of imperfect quantum efficiency, where we choose $\eta_1=\eta_2=0.8$. For comparison purposes, we show the left the same for the mod 4 implementation (also shown in Fig. \ref{fig_7}, bottom right panel). While for $\alpha\ll1$, both implementations perform similarly, for larger values of $\alpha$ when photon loss becomes more dominant, it is more advantageous to use the (mod 4) $+ \rm{P}_a + \rm{P}_b$ implementation over the mod 4 implementation. This is because unlike the mod 4 implementation which corrects for loss of single photons to first order, the (mod 4) $+ \rm{P}_a + \rm{P}_b$ implementation corrects for the loss of single photons in each of arnie and bert. 

\begin{figure}
\centering
\includegraphics[width = 0.5\textwidth]{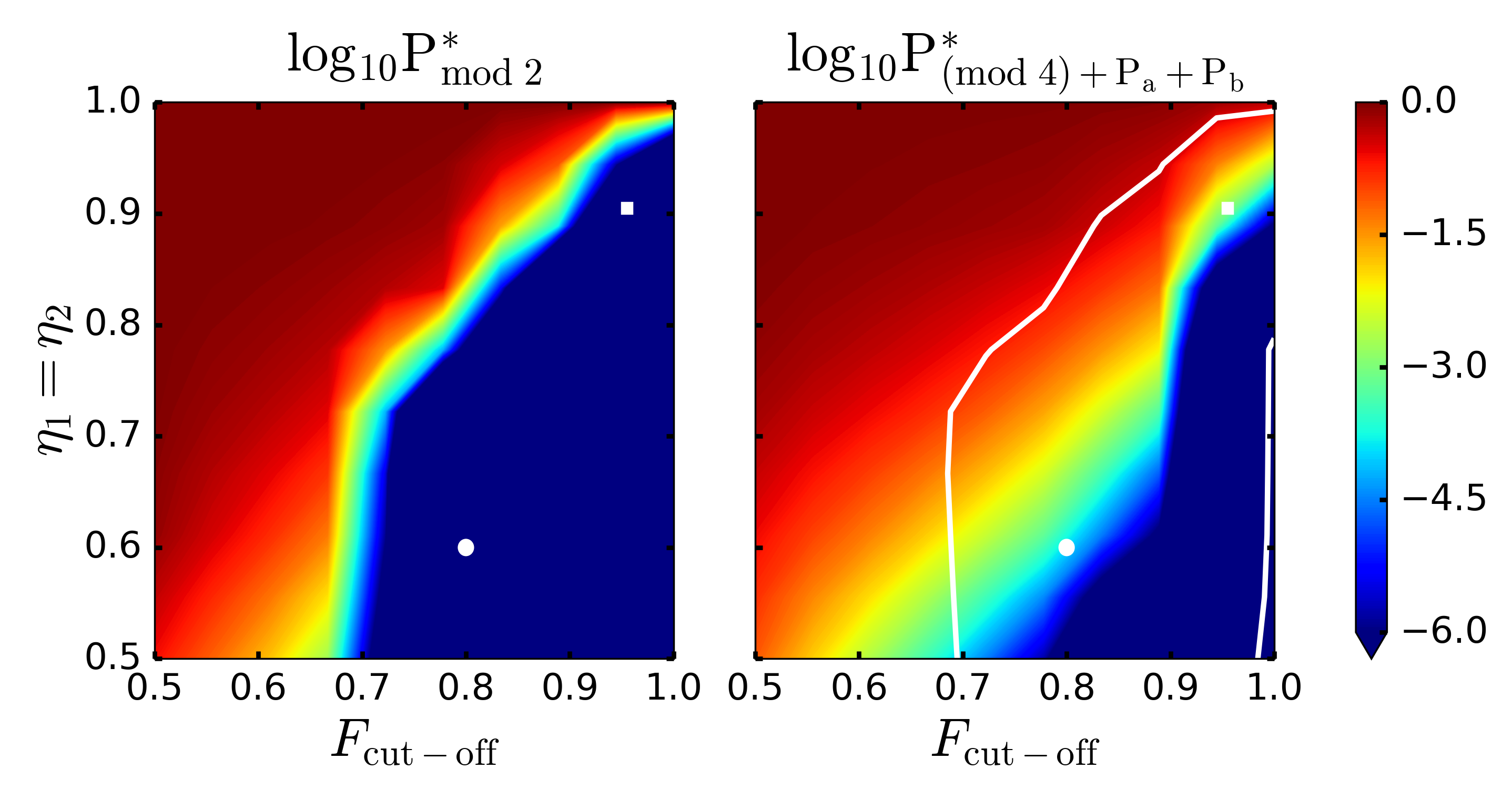}
\caption{ \label{fig_10} (color online) Optimized total success-probability $P^*$ for different cut-off fidelities and different choices of $\eta_1,\eta_2$ for the mod 2 (left panel) and the (mod 4) $+ \rm{P}_a + \rm{P}_b$ (right panel) implementations. The optimization is done for the value of the parameter $\alpha$. We choose, for simplicity, $\eta_1=\eta_2$. In contrast to the mod 2 implementation, the (mod 4) $+ \rm{P}_a + \rm{P}_b$ implementation shows substantially more resilience to lower efficiency. The white curves in the right panel enclose the region where  the (mod 4) $+ \rm{P}_a + \rm{P}_b$ implementation has a higher success rate than the mod 2 implementation. For instance, the (mod 4) $+ \rm{P}_a + \rm{P}_b$ implementation gives rise to Bell-states with fidelity $\geq0.95$ with a success rate of $10^{-2}$ for an efficiency of $\eta_1=\eta_2=0.9$ (the white rectangle in each plot). This should be compared with a success rate of less than $10^{-10}$ for the mod 2 implementation (left panel) and that of $10^{-4}$ for the mod 4 implementation (right panel of Fig. \ref{fig_8}). Even with efficiency values achievable in current circuit-QED systems of $\eta_1=\eta_2=0.6$, one can generate entangled states with overlaps to Bell-states $\geq0.8$ with a success-rate of $10^{-4}$ (the white circle in each plot). However, for low enough efficiencies, the error correcting protocol ceases to be advantageous since higher order photon loss become more important. For higher order protection, one will have to resort to different encodings \cite{Gottesman_Preskill_2001, Leghtas_Mirrahimi_2013, Michael_Girvin_2016, Terhal_Weigand_2016}. }
\end{figure}

As before, for each value of cut-off fidelity and inefficiency, an optimal choice of $\alpha$ can be obtained. This optimization is done numerically. The optimized success-rate as a function of inefficiency and cut-off fidelity is shown below (Fig.\ \ref{fig_10}). We again choose $\eta_1=\eta_2$ for simplicity. For comparison purposes, we show the optimized probability of success for the mod 2 implementation (what is shown also in Fig. \ref{fig_8}, left panel). The white curves in the right panel enclose the region where the (mod 4) $+ \rm{P}_a + \rm{P}_b$ implementation has a higher success rate than the mod 2 implementation. In particular,  for $\eta_1=\eta_2=0.9$, only the (mod 4) $+ \rm{P}_a + \rm{P}_b$ implementation is able to give rise to Bell-states with overlap $\geq0.95$ with a success-rate of $\sim10^{-2}$, whereas the mod 2 implementation has a success rate less than $10^{-10}$ (the white rectangles in the plots). Even with efficiency values achievable in current circuit-QED systems of $\eta_1=\eta_2=0.6$, with the (mod 4) $+ \rm{P}_a + \rm{P}_b$ implementation, one can generate entangled states with overlaps to Bell-states $\geq0.8$ with a success-rate of $10^{-4}$ (the white circles in the plots). Note that, this protocol is not able to generate perfect fidelity Bell-states for efficiency parameters of around $0.6$. This is because the mod 4 encoding protects against photon loss to first order. For higher order protection, one will have to resort to different encodings \cite{Gottesman_Preskill_2001, Leghtas_Mirrahimi_2013, Michael_Girvin_2016, Terhal_Weigand_2016}. 

\section{Conclusion}
\label{concl}
To summarize, we have presented in this paper, a protocol to remotely entangle two distant, mutually non-interacting, stationary qubits. To that end, we have used a propagating ancilla qubit for each of the stationary qubit. In the first step, local entanglement is generated between each stationary qubit and its associated ancilla. Subsequently, a joint two-qubit measurement is performed on the propagating ancilla qubits, followed by individual single-qubit measurements on the same. Depending on the three measurement outcomes, the two stationary qubits are projected on to an entangled state. We have discussed two continuous variable implementation of our protocol. In the first implementation, the ancilla qubits were encoded in even and odd Schr\"odinger cat states. For this encoding, the two-qubit measurement was done by a joint-photon-number-modulo-2 measurement and the single-qubit measurements were performed by homodyne detections. Subsequently, we described a second implementation, where the ancilla qubits were encoded in mod 4 cat states. For this encoding, the two-qubit measurement was performed by a joint-photon-number-modulo-4 measurement and the single qubit measurements were performed by homodyne detections. 
We analyzed the resilience of the two implementations to finite quantum efficiency arising out of  imperfections in realistic quantum systems. We described how with the mod 4 implementation, it is possible to  suppress loss of coherence due to loss of a photon in either of the ancilla qubits. Lastly, we presented an improvement of the mod 4 implementation, where we made individual photon-number-modulo-2 measurements of the ancilla qubits, together with the joint-photon-number-modulo-4 measurement, by virtue of which we suppressed the decoherence due to loss of a photon in both the ancilla qubits. We demonstrated that it is indeed possible to trade-off a higher success rate, present in the mod 2 implementation, for a higher fidelity of the generated entangled state, present in the (mod 4) $+ \rm{P}_a + \rm{P}_b$ implementation, using error correction. 

Next, we point out two future directions of research that this work leads to. First, the use of homodyne detection as the single-qubit measurement in the final step of the mod 4 or the (mod 4) $+ \rm{P}_a + \rm{P}_b$ implementation of our protocol lowers the success-rate of generating entangled states. This is because both arnie and bert are in superpositions of $|C_\alpha^+\rangle$ and $|C_{i\alpha}^+\rangle$, and thus, irrespective of the choice of the quadrature, the homodyne measurement is always made on the complementary quadrature of the modes for one of the cat states. This gives rise to fringes in the resultant overlap to the Bell-states and lowers the success rate of generating the same. It will be worthwhile to explore alternatives for the homodyne detection to boost the success rate of the error correcting protocol. Second, the error correcting encoding we used is designed to protect against losses of single-photons to first order. This is why mod 4 implementation protects against loss of a photon in either of the ancilla qubits. By including individual parity measurements, in addition to the joint-photon-number-modulo-4 measurement, we corrected for the decoherence due to loss of a photon in both of the ancillas. However, with this encoding, higher order photon loss errors cannot be corrected. It will be interesting to explore different encodings of the ancilla qubits for protection against higher order photon loss \cite{Gottesman_Preskill_2001, Leghtas_Mirrahimi_2013, Michael_Girvin_2016, Terhal_Weigand_2016}. 

Discussions with Michel Devoret, Michael Hatridge, Shyam Shankar, Matti Silveri, Steve Girvin, Chen Wang, Barbara Terhal and   Rob Schoelkopf are gratefully acknowledged. The work was supported by US Army Research Office Grant No. W911NF-14-1-0011 and NSF grant ECCS 1068642. LJ acknowledges the support from ARL-CDQI, ARO W911NF-14-1- 0563, AFOSR MURI FA9550-14-1-0052, FA9550-14-1-0015, Alfred P. Sloan Foundation BR2013-049, and the Packard Foundation
2013-39273.

\appendix

\section{Protocol to realize a joint-photon-number-modulo-4 measurement}
\label{mod 4-impl}
Here, we describe the protocol to perform the joint-photon-number-modulo-4 measurement of two resonator modes. Consider two resonator modes (with annihilation operators $ \mathbf{a},  \mathbf{b}$), which are dispersively coupled to a transmon qubit (whose ground, excited states are denoted by $|g\rangle, |e\rangle$). We require the dispersive coupling strength to be equal \footnote{Equal coupling of the transmon qubit to two different cavity modes is challenging to realize experimentally and is not a pre-requisite for making these joint-photon-number measurements. It can be avoided by using higher excited states of the transmon as was demonstrated in the joint-photon-number-modulo-2 measurements of \cite{Wang_Schoelkopf_2016}.} for each of the modes $\mathbf{a},\mathbf{b}$.
The resultant Hamiltonian describing the two cavity modes and the transmon qubit is given by: 
\begin{eqnarray}
\mathbf{H}_{\rm{mod 4}} &=& \omega_q|e\rangle\langle e| + \omega_a\mathbf{a}^\dagger\mathbf{a} + \omega_b\mathbf{b}^\dagger\mathbf{b}\nonumber\\&&-\chi(\mathbf{a}^\dagger\mathbf{a}+\mathbf{b}^\dagger\mathbf{b})|e\rangle\langle e|,
\end{eqnarray}
where $\chi$ is the cross-Kerr coupling of the transmon qubit to the cavity modes. A joint-photon-number-modulo-4 measurement can be performed in the following way. First, a joint-photon-number-modulo-2 measurement is performed \cite{Wang_Schoelkopf_2016}. This can be done by exciting the transmon qubit at frequencies $\omega_q-2k\chi$ where $k\in\mathbb{Z}$, followed by a Z measurement of the transmon. A Z measurement result of $\tilde{p}_1=1(-1)$ corresponds to a joint-photon-number of the modes $\mathbf{a},\mathbf{b}$ being $2k(2k+1)$. Second, a measurement is performed that reveals if the joint-photon-number of the arnie and bert modes $\in\{4k,4k+1\}$ or not, where $k\in\mathbb{Z}$. This can be done by using the procedure as making the joint-photon-number-modulo-2 measurement. The only difference is that the transmon qubit is now excited with frequencies $\omega_q-4k\chi, \omega_q-(4k+1)\chi$, $k\in\mathbb{Z}$. In this case, a Z measurement outcome of $\tilde{p}_2 = 1$ corresponds to the joint-photon-number of the modes $\mathbf{a},\mathbf{b}$ $\in\{4k,4k+1\}$, while $\tilde{p}_2 = -1$ corresponds to the same $\in\{4k+2,4k+3\}$. From these two measurement outcomes $\tilde{p}_1,\tilde{p}_2$, we can infer the joint-photon-number-modulo-4 outcome. For instance, $\tilde{p}_1=\tilde{p}_2 = 1$, the joint-photon-numer-modulo-4 outcome $\lambda = 0$. Similarly, $\tilde{p}_1=-1,\tilde{p}_2 = 1\Rightarrow\lambda = 1$, $\tilde{p}_1=1,\tilde{p}_2 = -1\Rightarrow\lambda = 2$ and $\tilde{p}_1=-1,\tilde{p}_2 = -1\Rightarrow\lambda = 3$. Obviously, in an actual experiment, one doesn't need to send an infinite set of frequencies to make these measurements. The actual number of frequencies depend on the photon-number distributions of the two resonator modes. 


\section{Computation of probability of outcomes and overlap to the Bell-states in absence of imperfections}
\label{AppA}
\subsection{Implementation using Schr\"odinger cat states}
\label{appAmod2}
In this section, we outline the computation of the probability of success and overlaps to the different Bell-states. To that end, we start with the state of the four modes: Alice, Bob, arnie and bert: 
\begin{eqnarray}
|\Psi^{p=1}_{\rm{ABab}}\rangle &=& \frac{1}{\sqrt{2}}\big(|g,g,C_\alpha^+,C_\alpha^+\rangle + |e,e,C_\alpha^-,C_\alpha^-\rangle\big),\\|\Psi^{p=-1}_{\rm{ABab}}\rangle &=& \frac{1}{\sqrt{2}}\big(|g,e,C_\alpha^+,C_\alpha^-\rangle + |e,g,C_\alpha^-,C_\alpha^+\rangle\big).
\end{eqnarray}
The homodyne detection can be modeled as a projection of arnie and bert on $x$ eigenstates, described by the projection operator: ${\cal M}_X=|x_a,x_b\rangle\langle x_a,x_b|$. Consider the case $p=1$. After the homodyne detection, the un-normalized wave-function for the modes of Alice, Bob, arnie and bert is given by: 
\begin{eqnarray}
{\cal M}_X|\Psi^{p=1}_{\rm{ABab}}\rangle&=&\frac{1}{\sqrt{2}}\big(\langle x_a|C_\alpha^+\rangle\langle x_b|C_\alpha^+\rangle|g,g\rangle\nonumber\\&& + \langle x_a|C_\alpha^-\rangle\langle x_b|C_\alpha^-\rangle|e,e\rangle\big)|x_a,x_b\rangle.
\end{eqnarray}
Using the wave-function in the position basis of an even (odd) Schr\"odinger cat state: 
\begin{eqnarray}
\label{homodyne_even_odd_cat}
\langle x|C_\alpha^\pm\rangle = \Big(\frac{2}{\pi}\Big)^{1/4}{\cal N}_\pm e^{-x^2-\alpha^2}\big(e^{2x\alpha}\pm e^{-2x\alpha}\big)\nonumber,
\end{eqnarray}
we arrive at:
\begin{eqnarray}
{\cal M}_X|\Psi^{p=1}_{\rm{ABab}}\rangle&=&\frac{2}{\sqrt{\pi}}e^{-(x_a^2+x_b^2)}e^{-2\alpha^2}\Big[\frac{\cosh(2x_a\alpha)\cosh(2x_b\alpha)}{1+e^{-2\alpha^2}}\nonumber\\&&|g,g\rangle+\frac{\sinh(2x_a\alpha)\sinh(2x_b\alpha)}{1-e^{-2\alpha^2}}|e,e\rangle\Big]|x_a,x_b\rangle\nonumber.
\end{eqnarray}
The probability distribution of outcomes $P(x_a,x_b)$ and the resultant state of Alice and Bob $\rho_{\rm{AB}}$ are then given by:
\begin{eqnarray}
P^{p=1}(x_a,x_b)&=&\frac{1}{2}{\rm{Tr}}\Big[{\cal M}_X|\Psi^{p=1}_{\rm{ABab}}\rangle\langle\Psi^{p=1}_{\rm{ABab}}|{\cal M}_X^\dagger\Big],\\
\rho^{p=1}_{\rm{AB}}&=&\frac{{\rm{Tr}}_{\rm{ab}}\Big[{\cal M}_X|\Psi^{p=1}_{\rm{ABab}}\rangle\langle\Psi^{p=1}_{\rm{ABab}}|{\cal M}_X^\dagger\Big]}{{\rm{Tr}}\Big[{\cal M}_X|\Psi^{p=1}_{\rm{ABab}}\rangle\langle\Psi^{p=1}_{\rm{ABab}}|{\cal M}_X^\dagger\Big]},
\end{eqnarray}
giving rise to: Eqs.\ \eqref{prob_dist_mod2_perf}, \eqref{overlap_mod2_perf}, \eqref{overlap_mod2_perf_1}. Note the factor of $1/2$, which arises in the total probability of outcomes due to the fact that the $p=1$ outcome happens with a probability $1/2$. The calculation for $p=-1$ can be done in an analogous fashion. 

\subsection{Implementation using mod 4 cat states}
\label{appAmod4}
In this section, we outline the calculation for the probability of success and the overlap to the Bell-states when the ancilla qubits are encoded in mod 4 cat states. The state of Alice, Bob, arnie and Bert, following the joint-photon-number-modulo-4 measurement can be written as [Eq.\ \eqref{lambda=0_eqn}, \eqref{lambda=2_eqn}]:  
\begin{eqnarray}
|\Psi^{\lambda=0}_{\rm{ABab}}\rangle &=& \frac{1}{\sqrt{2}}\big(|g,g,C_\alpha^{0{\rm{mod}}4},C_\alpha^{0{\rm{mod}}4}\rangle\nonumber\\&& + |e,e,C_\alpha^{2{\rm{mod}}4},C_\alpha^{2{\rm{mod}}4}\rangle\big)\nonumber\\|\Psi^{\lambda=2}_{\rm{ABab}}\rangle &=& \frac{1}{\sqrt{2}}\big(|g,e,C_\alpha^{0{\rm{mod}}4},C_\alpha^{2{\rm{mod}}4}\rangle\nonumber\\&& + |e,g,C_\alpha^{2{\rm{mod}}4},C_\alpha^{0{\rm{mod}}4}\rangle\big).
\end{eqnarray}
To compute resultant states of Alice and Bob after the subsequent homodyne detection of arnie and bert, we use the following definitions of the mod 4 cats: 
\begin{equation}
\label{four_cat_defn_1}
|C_{\alpha}^{\lambda{\rm{mod}}4}\rangle = \tilde{\cal N}_{\lambda}\Big(\Big|C_{\alpha}^{(-)^\lambda}\Big\rangle+(-i)^\lambda\Big|C_{i\alpha}^{(-)^\lambda}\Big\rangle\Big),
\end{equation}
where $\tilde{\cal N}_\lambda = \big[2+2(-i)^\lambda\{e^{i|\alpha|^2}+(-1)^\lambda e^{-i|\alpha|^2}\}/
\{e^{|\alpha|^2}+(-1)^\lambda e^{-|\alpha|^2}\}\big]^{-\frac{1}{2}}$, $\lambda\in\{0,1,2,3\}$ and $(-)^\lambda = +(-)$ for even (odd) $\lambda$. This leads to:
\begin{eqnarray}
\langle x|C_{\alpha}^{\lambda{\rm{mod}}4}\rangle = \tilde{\cal N}_\lambda\big(\langle x|C_{\alpha}^{+}\rangle+i^\lambda\langle x|C_{i\alpha}^{+}\rangle\big), \ \lambda=0,2\nonumber
\end{eqnarray}
where $\langle x|C_{\alpha}^+\rangle$ and $\langle x|C_{i\alpha}^+\rangle$ are given by:
\begin{eqnarray}
\langle x|C_{\alpha}^+\rangle &=& 2\Big(\frac{2}{\pi}\Big)^{1/4}{\cal N}_+e^{-x^2-\alpha^2}\cosh(2\alpha x),\\
\langle x|C_{i\alpha}^+\rangle &=& 2\Big(\frac{2}{\pi}\Big)^{1/4}{\cal N}_+e^{-x^2}\cos(2\alpha x).
\end{eqnarray}
Thus, we arrive at:
\begin{eqnarray}
{\cal M}_X|\Psi^{\lambda=0}_{\rm{ABab}}\rangle &=& \frac{1}{\sqrt{2}}\big(\langle x_a,x_b|C_{\alpha}^{0{\rm{mod}}4},C_{\alpha}^{0{\rm{mod}}4}\rangle|g,g\rangle\nonumber\\&& + \langle x_a,x_b|C_{\alpha}^{2{\rm{mod}}4},C_{\alpha}^{2{\rm{mod}}4}\rangle|e,e\rangle\big)|x_a,x_b\rangle\nonumber\\&=&\frac{4}{\sqrt{\pi}}{\cal N}_+^2e^{-(x_a^2+x_b^2)}\Big[\tilde{\cal N}_0^2F_0(x_a)F_0(x_b)|g,g\rangle\nonumber\\&& + \tilde{\cal N}_2^2F_2(x_a)F_2(x_b)|e,e\rangle\Big]|x_a,x_b\rangle,
\end{eqnarray}
where $F_\lambda(x) = e^{-\alpha^2}\cosh(2\alpha x)+i^\lambda\cos(2\alpha x)$.
This leads to the probability distribution and the resultant density matrix for Alice and Bob through the relations: 
\begin{eqnarray}
P^{\lambda=0}(x_a,x_b)&=&\frac{1}{2}{\rm{Tr}}\Big[{\cal M}_X|\Psi^{\lambda=0}_{\rm{ABab}}\rangle\langle\Psi^{\lambda=0}_{\rm{ABab}}|{\cal M}_X^\dagger\Big],\\
\rho_{\rm{AB}}^{\lambda=0}&=&\frac{{\rm{Tr}}_{\rm{ab}}\Big[{\cal M}_X|\Psi^{\lambda=0}\rangle\langle\Psi^{\lambda=0}_{\rm{ABab}}|{\cal M}_X^\dagger\Big]}{{\rm{Tr}}\Big[{\cal M}_X|\Psi^{\lambda=0}_{\rm{ABab}}\rangle\langle\Psi^{\lambda=0}|{\cal M}_X^\dagger\Big]}.
\end{eqnarray}
Similar set of calculations can be done for the outcome $\lambda =2$.
\section{Computation of propagating qubit-photon states in presence of imperfections}
\label{AppA_1}
\subsection{Implementation using Schr\"odinger cat states}
\label{appAmod2_1}
First, we describe the computation of the state of Alice and arnie after their entangled qubit-photon states decohere as they propagate through the transmission line. The initial state of Alice and arnie is given by [Eq.\ \eqref{Aa_init_eq}]: 
\begin{equation}
|\Psi_{\rm{Aa}}\rangle =\frac{1}{\sqrt{2}}\sum_{j,\mu=0}^1{\cal N}_j(-1)^{j\mu}|j,(-1)^\mu\alpha\rangle.
\end{equation}
To compute the final state after attenuation losses, first we introduce an auxiliary mode a$'$, initialized to vacuum. Then, we model the losses by passing the joint system of Alice, arnie and a$'$ through a beam-splitter with transmission probability $\eta_1$. 
Thus, the state evolves according to: 
\begin{eqnarray}
|\Psi_{\rm{Aa}}\rangle&\otimes&|0\rangle \rightarrow |\bar\Psi_{\rm{Aa}}\rangle\nonumber\\|\bar\Psi_{\rm{Aa}}\rangle&=&\frac{1}{\sqrt{2}}\sum_{j,\mu=0}^1{\cal N}_j(-1)^{j\mu}|j,(-1)^\mu\bar\alpha,(-1)^\mu\epsilon\rangle, 
\end{eqnarray}
where $\bar\alpha=\sqrt{\eta_1}\alpha$ and $\epsilon = \sqrt{1-\eta_1}\alpha$. 
Thus, the density matrix for the modes Alice, arnie and a$'$ can be written as: 
\begin{eqnarray}
\bar\rho_{\rm{Aa}} &=& \frac{1}{2}\sum\limits_{j,j',\mu,\mu'=0}^1{\cal N}_j{\cal N}_{j'}(-1)^{\bm{\mu}\cdot\bm{j}}\big(\big|j,(-1)^\mu\bar\alpha\big\rangle\nonumber\\&&\big\langle j',(-1)^{\mu'}\bar\alpha\big|\big)|(-1)^\mu\epsilon\rangle\langle (-1)^{\mu'}\epsilon|\nonumber\\\label{Aa_eqn_int}\Rightarrow \rho_{\rm{Aa}} &=& \frac{1}{2}\sum\limits_{j,j',\mu,\mu'=0}^1{\cal N}_j{\cal N}_{j'}(-1)^{\bm{\mu}\cdot\bm{j}}e^{-\epsilon^2\{1-(-1)^{\mu+\mu'}\}}\nonumber\\&&\big(\big|j,(-1)^\mu\bar\alpha\big\rangle\big\langle j',(-1)^{\mu'}\bar\alpha\big|\big)
\end{eqnarray}
where $\bm{\mu} = \{\mu,\mu'\}, \bm{j} = \{j,j'\}$ and in the last line, we have traced out the auxiliary mode a$'$. Here, $e^{-\epsilon^2\{1-(-1)^{\mu+\mu'}\}}$ shows explicitly the loss of coherence due to the loss of information to the environment. At this point, we need to re-express the $|(-1)^\mu\bar\alpha\rangle$ in the eigenbasis of the measurement operator: joint-photon-number-modulo-2. To that end, we use: 
\begin{eqnarray}
\label{coh_cat_eq}
|(-1)^\mu\bar\alpha\rangle = \frac{1}{2}\sum\limits_{k=0}^1\frac{(-1)^{\mu k}}{\bar{\cal N}_k}\big|C_{\bar\alpha}^{(-)^k}\big\rangle,
\end{eqnarray}
where $\bar{\cal N}_j = 1/\sqrt{2(1+(-1)^j e^{-2|\bar\alpha|^2})}$.
Using Eqs.\ \eqref{Aa_eqn_int}, \eqref{coh_cat_eq}, we arrive at the density matrix for Alice and arnie: 
\begin{eqnarray}
\rho_{\rm{Aa}} &=& \frac{1}{8}\sum\limits_{\rm{Aa}}\frac{{\cal N}_j{\cal N}_{j'}}{\bar{\cal N}_k\bar{\cal N}_{k'}}(-1)^{\bm{\mu}\cdot(\bm{j}+\bm{k})}e^{-\epsilon^2\{1-(-1)^{\mu+\mu'}\}}\nonumber\\&&\Big|j,C_{\bar\alpha}^{(-)^k}\Big\rangle\Big\langle j',C_{\bar\alpha}^{(-)^{k'}}\Big|,
\end{eqnarray}
where $\sum\limits_{\rm{Aa}} = \sum\limits_{j,j'=0}^1\sum\limits_{k,k'=0}^1\sum\limits_{\mu,\mu'=0}^1, \bm{k} = \{k,k'\}$. 

Similar calculations can be done for the entangled states of Bob and bert, yielding:
\begin{eqnarray}
\rho_{\rm{Bb}} &=& \frac{1}{8}\sum\limits_{\rm{Bb}}\frac{{\cal N}_l{\cal N}_{l'}}{\bar{\cal N}_m\bar{\cal N}_{m'}}(-1)^{\bm{\nu}\cdot(\bm{l}+\bm{m})}e^{-\epsilon^2\{1-(-1)^{\nu+\nu'}\}}\nonumber\\&&\Big|l,C_{\bar\alpha}^{(-)^m}\Big\rangle\Big\langle l',C_{\bar\alpha}^{(-)^{m'}}\Big|,
\end{eqnarray}
where $\sum\limits_{\rm{Bb}} = \sum\limits_{l,l'=0}^1\sum\limits_{m,m'=0}^1\sum\limits_{\nu,\nu'=0}^1, \bm{\nu} = \{\nu,\nu'\}, \bm{l} = \{l,l'\}$ and $\bm{m} = \{m,m'\}$.
The tensor product of $\rho_{\rm{Aa}}$ and $\rho_{\rm{Bb}}$ gives Eq.\ \eqref{rhoABab_eqn}.
\subsection{Implementation using mod 4 cat states}
\label{appAmod4_1}
First, we describe the computation of the qubit-photon states of Alice and arnie after propagation through the transmission line. The starting point is Eq.\ \eqref{Aa_init_eq_mod4}:
\begin{eqnarray}
|\Psi_{\rm{Aa}}\rangle &=&\frac{1}{\sqrt{2}}\sum_{j,\mu,\nu=0}^1\tilde{\cal N}_{2j}{\cal N}_{2j}(-1)^{j\nu}|j,(-1)^\mu i^\nu\alpha\rangle.
\end{eqnarray}
To compute the entangled qubit-photon state when it arrives at the joint-photon-number-modulo-4 measurement apparatus, we use the approach outlined in Appendix. \ref{appAmod2_1}, introducing an auxiliary mode a$'$ in vacuum, computing the resultant state of Alice, arnie and a$'$ as it passes through a beam-splitter of transmission probability $\eta_1$ and subsequently, tracing out the mode a$'$. Following the notation in Appendix. \ref{appAmod2_1}, we find that:
\begin{eqnarray}
|\bar\Psi_{\rm{Aa}}\rangle&=&\frac{1}{\sqrt{2}}\sum_{j,\mu,\nu=0}^1\tilde{\cal N}_{2j}{\cal N}_{2j}(-1)^{j\nu}|j,(-1)^\mu i^\nu\bar\alpha,(-1)^\mu i^\nu\epsilon\rangle\nonumber, 
\end{eqnarray}
where $\bar\alpha=\sqrt{\eta_1}\alpha$ and $\epsilon = \sqrt{1-\eta_1}\alpha$. 
Thus, the density matrix for the modes Alice, arnie and a$'$ can be written as: 
\begin{eqnarray}
\bar\rho_{\rm{Aa}} &=& \frac{1}{2}\sum\limits_{\{\}}\tilde{\cal N}_{2j}\tilde{\cal N}_{2j'}{\cal N}_{2j}{\cal N}_{2j'}(-1)^{\bm{\nu}\cdot\bm{j}}\big(\big|j,(-1)^\mu i^\nu\bar\alpha\big\rangle\nonumber\\&&\big\langle j',(-1)^{\mu'}i^{\nu'}\bar\alpha\big|\big)|(-1)^\mu i^\nu\epsilon\rangle\langle (-1)^{\mu'}i^{\nu'}\epsilon|\nonumber\\\label{Aa_eqn_int_mod4}\Rightarrow \rho_{\rm{Aa}} &=& \frac{1}{2}\sum\limits_{\{\}}\tilde{\cal N}_{2j}\tilde{\cal N}_{2j'}{\cal N}_{2j}{\cal N}_{2j'}(-1)^{\bm{\nu}\cdot\bm{j}}\nonumber\\&&e^{-\epsilon^2\{1-(-1)^{\mu+\mu'}i^{\nu-\nu'}\}}\big|j,(-1)^\mu i^\nu\bar\alpha\big\rangle\big\langle j',(-1)^{\mu'}i^{\nu'}\bar\alpha\big|\nonumber,
\end{eqnarray}
where $\sum\limits_{\{\}}=\sum\limits_{j,j'=0}^1 \sum\limits_{\mu,\mu'=0}^1 \sum\limits_{\nu,\nu'=0}^1$, $\bm{\nu} = \{\nu,\nu'\}, \bm{j} = \{j,j'\}$ and in the last line, we have traced out the auxiliary mode a$'$. In the next step, we express the above equation in the eigenbasis of the joint-photon-number-modulo-4 measurement. To that end, we use: 
\begin{eqnarray}
\big|(-1)^\mu i^\nu\bar\alpha\big\rangle &=& \frac{1}{4}\sum\limits_{\gamma=0}^3\frac{1}{\bar{\tilde{\cal N}}_\gamma\bar{\cal N}_\gamma}\big|C_{(-1)^\mu i^\nu\bar\alpha}^{\gamma{\rm{mod}}4}\big\rangle\nonumber\\&=&\frac{1}{4}\sum\limits_{\gamma=0}^3\frac{(-1)^{\mu\gamma}i^{\nu\gamma}}{\bar{\tilde{\cal N}}_\gamma\bar{\cal N}_\gamma}\big|C_{\bar\alpha}^{\gamma{\rm{mod}}4}\big\rangle,
\end{eqnarray}
where $\bar{\tilde{\cal N}}_\gamma,\bar{\cal N}_\gamma$ can obtained from the definitions of ${\tilde{\cal N}}_\gamma,{\cal N}_\gamma$ (cf. Secs. \ref{mod 2_imperf}, \ref{mod 4_imperf}) by making the substitution $\alpha\rightarrow\bar\alpha$ and the last line follows from the definition of mod 4 cats (see Sec. 2.2 of \cite{Mirrahimi_Devoret_2014}). Combining the last two equations results in: 
\begin{eqnarray}
\rho_{\rm{Aa}}& =& \frac{1}{2^5}\sum\limits_{\rm{Aa}}\frac{\tilde{\cal N}_{2j}\tilde{\cal N}_{2j'}{\cal N}_{2j}{\cal N}_{2j'}}{\bar{\tilde{\cal N}}_{\gamma}\bar{\tilde{\cal N}}_{\gamma'}\bar{\cal N}_{\gamma}\bar{\cal N}_{\gamma'}}(-1)^{\bm{\nu}\cdot\bm{j}+\bm{\mu}\cdot\bm{\gamma}}i^{\nu\gamma-\nu'\gamma'}\nonumber\\&&e^{-\epsilon^2\{1-(-1)^{\mu+\mu'}i^{\nu-\nu'}\}}\Big|j,C_{\bar\alpha}^{\gamma{\rm{mod}}4}\Big\rangle\Big\langle j',C_{\bar\alpha}^{\gamma'{\rm{mod}}4}\Big|,\nonumber
\end{eqnarray}
where $\sum\limits_{\rm{Aa}} = \sum\limits_{j,j'=0}^1 \sum\limits_{\mu,\mu'=0}^1 \sum\limits_{\nu,\nu'=0}^1\sum\limits_{\gamma,\gamma'=0}^3$. Here, we have used, as before, the following definitions: $\bm{j} = \{j,j'\}, \bm{\mu} = \{\mu,\mu'\}, \bm{\nu}=\{\nu,\nu'\}$ and $\bm{\gamma} = \{\gamma, \gamma'\}$. Similar calculations can be done for Bob and bert, yielding:
\begin{eqnarray}
\rho_{\rm{Bb}}& =& \frac{1}{2^5}\sum\limits_{\rm{Bb}}\frac{\tilde{\cal N}_{2k}\tilde{\cal N}_{2k'}{\cal N}_{2k}{\cal N}_{2k'}}{\bar{\tilde{\cal N}}_{\delta}\bar{\tilde{\cal N}}_{\delta'}\bar{\cal N}_{\delta}\bar{\cal N}_{\delta'}}(-1)^{\bm{\psi}\cdot\bm{k}+\bm{\phi}\cdot\bm{\delta}}i^{\psi\delta-\psi'\delta'}\nonumber\\&&e^{-\epsilon^2\{1-(-1)^{\phi+\phi'}i^{\psi-\psi'}\}}\Big|j,C_{\bar\alpha}^{\delta{\rm{mod}}4}\Big\rangle\Big\langle j',C_{\bar\alpha}^{\delta'{\rm{mod}}4}\Big|,\nonumber
\end{eqnarray}
where $\sum\limits_{\rm{Bb}} = \sum\limits_{k,k'=0}^1 \sum\limits_{\phi,\phi'=0}^1 \sum\limits_{\psi,\psi'=0}^1\sum\limits_{\delta,\delta'=0}^3$ and $\bm{k} = \{k,k'\}, \bm{\phi} = \{\phi,\phi'\}, \bm{\psi}=\{\psi,\psi'\}$ and $\bm{\delta} = \{\delta, \delta'\}$. The tensor product of $\rho_{\rm{Aa}}$ and $\rho_{\rm{Bb}}$ gives us Eq.\ \eqref{ABab_mod_4_imperf}. 

\subsection{Probability distribution and overlaps to Bell-states for $\lambda=1,3$}
\label{lambda_1_3}
\begin{figure}[!h]
\centering
\includegraphics[width = 0.5\textwidth]{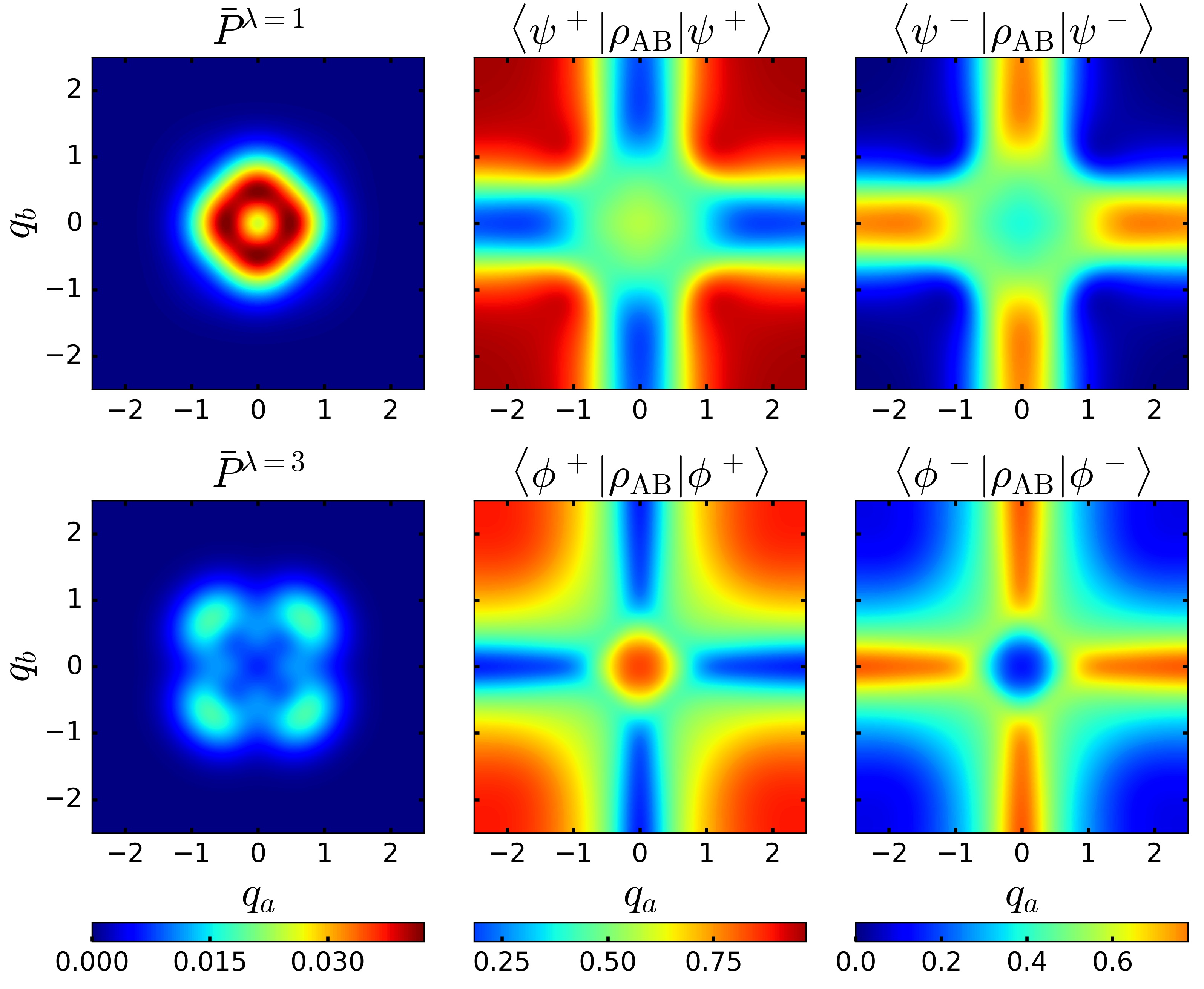}
\caption{ \label{fig_6_a} (color online) Probability distribution $\bar{P}^\lambda(q_a, q_b)$ of outcomes of the homodyne measurements of arnie and bert and resulting overlap of Alice and Bob's joint density matrix $\rho_{\rm{AB}}$ with the Bell-states $|\phi^\pm\rangle = (|g,g\rangle \pm|e,e\rangle)/\sqrt{2}, |\psi^\pm\rangle = (|g,e\rangle \pm|e,g\rangle)/\sqrt{2}$ is shown. We choose $\alpha=1$ and $\eta_1 = \eta_2=0.8$ and show the cases $\lambda=1,3$ (see Appendix. \ref{appAmod4_1} for $\lambda=1,3$). The top (bottom) left panel shows the probability of outcomes for the joint-photon-number-modulo-4 outcome to be $\lambda=1(3)$. Corresponding overlaps to the Bell-states $|\psi^\pm\rangle(|\phi^\pm\rangle)$ are plotted in the top (bottom) center and top (bottom) right panels. The overlaps to the even (odd) Bell-states for $\lambda=1(3)$ are not shown for brevity. }
\end{figure}

After the homodyne detection of arnie and bert, corresponding to the joint-photon-number-modulo-4 measurement outcomes $\lambda = 1,3$, the probability of success and the overlap to the Bell-states $|\phi^\pm\rangle, |\psi^\pm\rangle$ is shown in Fig.\ \ref{fig_6_a}. 

%

\bibliography{library}

\begin{thebibliography}{40}%
\makeatletter
\providecommand \@ifxundefined [1]{%
 \@ifx{#1\undefined}
}%
\providecommand \@ifnum [1]{%
 \ifnum #1\expandafter \@firstoftwo
 \else \expandafter \@secondoftwo
 \fi
}%
\providecommand \@ifx [1]{%
 \ifx #1\expandafter \@firstoftwo
 \else \expandafter \@secondoftwo
 \fi
}%
\providecommand \natexlab [1]{#1}%
\providecommand \enquote  [1]{``#1''}%
\providecommand \bibnamefont  [1]{#1}%
\providecommand \bibfnamefont [1]{#1}%
\providecommand \citenamefont [1]{#1}%
\providecommand \href@noop [0]{\@secondoftwo}%
\providecommand \href [0]{\begingroup \@sanitize@url \@href}%
\providecommand \@href[1]{\@@startlink{#1}\@@href}%
\providecommand \@@href[1]{\endgroup#1\@@endlink}%
\providecommand \@sanitize@url [0]{\catcode `\\12\catcode `\$12\catcode
  `\&12\catcode `\#12\catcode `\^12\catcode `\_12\catcode `\%12\relax}%
\providecommand \@@startlink[1]{}%
\providecommand \@@endlink[0]{}%
\providecommand \url  [0]{\begingroup\@sanitize@url \@url }%
\providecommand \@url [1]{\endgroup\@href {#1}{\urlprefix }}%
\providecommand \urlprefix  [0]{URL }%
\providecommand \Eprint [0]{\href }%
\providecommand \doibase [0]{http://dx.doi.org/}%
\providecommand \selectlanguage [0]{\@gobble}%
\providecommand \bibinfo  [0]{\@secondoftwo}%
\providecommand \bibfield  [0]{\@secondoftwo}%
\providecommand \translation [1]{[#1]}%
\providecommand \BibitemOpen [0]{}%
\providecommand \bibitemStop [0]{}%
\providecommand \bibitemNoStop [0]{.\EOS\space}%
\providecommand \EOS [0]{\spacefactor3000\relax}%
\providecommand \BibitemShut  [1]{\csname bibitem#1\endcsname}%
\let\auto@bib@innerbib\@empty
\bibitem [{\citenamefont {Ekert}(1991)}]{Ekert_1991}%
  \BibitemOpen
  \bibfield  {author} {\bibinfo {author} {\bibfnamefont {A.~K.}\ \bibnamefont
  {Ekert}},\ }\href {\doibase 10.1103/PhysRevLett.67.661} {\bibfield  {journal}
  {\bibinfo  {journal} {Phys. Rev. Lett.}\ }\textbf {\bibinfo {volume} {67}},\
  \bibinfo {pages} {661} (\bibinfo {year} {1991})}\BibitemShut {NoStop}%
\bibitem [{\citenamefont {Bennett}\ \emph {et~al.}(1993)\citenamefont
  {Bennett}, \citenamefont {Brassard}, \citenamefont {Cr{\'{e}}peau},
  \citenamefont {Jozsa}, \citenamefont {Peres},\ and\ \citenamefont
  {Wootters}}]{Bennett_Wootters_1993}%
  \BibitemOpen
  \bibfield  {author} {\bibinfo {author} {\bibfnamefont {C.~H.}\ \bibnamefont
  {Bennett}}, \bibinfo {author} {\bibfnamefont {G.}~\bibnamefont {Brassard}},
  \bibinfo {author} {\bibfnamefont {C.}~\bibnamefont {Cr{\'{e}}peau}}, \bibinfo
  {author} {\bibfnamefont {R.}~\bibnamefont {Jozsa}}, \bibinfo {author}
  {\bibfnamefont {A.}~\bibnamefont {Peres}}, \ and\ \bibinfo {author}
  {\bibfnamefont {W.~K.}\ \bibnamefont {Wootters}},\ }\href {\doibase
  10.1103/PhysRevLett.70.1895} {\bibfield  {journal} {\bibinfo  {journal}
  {Phys. Rev. Lett.}\ }\textbf {\bibinfo {volume} {70}},\ \bibinfo {pages}
  {1895} (\bibinfo {year} {1993})}\BibitemShut {NoStop}%
\bibitem [{\citenamefont {van Enk}\ \emph {et~al.}(1997)\citenamefont {van
  Enk}, \citenamefont {Cirac},\ and\ \citenamefont
  {Zoller}}]{van_Enk_Zoller_1997}%
  \BibitemOpen
  \bibfield  {author} {\bibinfo {author} {\bibfnamefont {S.~J.}\ \bibnamefont
  {van Enk}}, \bibinfo {author} {\bibfnamefont {J.~I.}\ \bibnamefont {Cirac}},
  \ and\ \bibinfo {author} {\bibfnamefont {P.}~\bibnamefont {Zoller}},\ }\href
  {\doibase 10.1103/PhysRevLett.78.4293} {\bibfield  {journal} {\bibinfo
  {journal} {Phys. Rev. Lett.}\ }\textbf {\bibinfo {volume} {78}},\ \bibinfo
  {pages} {4293} (\bibinfo {year} {1997})}\BibitemShut {NoStop}%
\bibitem [{\citenamefont {Briegel}\ \emph {et~al.}(1998)\citenamefont
  {Briegel}, \citenamefont {D{\"{u}}r}, \citenamefont {Cirac},\ and\
  \citenamefont {Zoller}}]{Briegel_Zoller_1998}%
  \BibitemOpen
  \bibfield  {author} {\bibinfo {author} {\bibfnamefont {H.-J.}\ \bibnamefont
  {Briegel}}, \bibinfo {author} {\bibfnamefont {W.}~\bibnamefont {D{\"{u}}r}},
  \bibinfo {author} {\bibfnamefont {J.~I.}\ \bibnamefont {Cirac}}, \ and\
  \bibinfo {author} {\bibfnamefont {P.}~\bibnamefont {Zoller}},\ }\href
  {\doibase 10.1103/PhysRevLett.81.5932} {\bibfield  {journal} {\bibinfo
  {journal} {Phys. Rev. Lett.}\ }\textbf {\bibinfo {volume} {81}},\ \bibinfo
  {pages} {5932} (\bibinfo {year} {1998})}\BibitemShut {NoStop}%
\bibitem [{\citenamefont {Clauser}\ \emph {et~al.}(1969)\citenamefont
  {Clauser}, \citenamefont {Horne}, \citenamefont {Shimony},\ and\
  \citenamefont {Holt}}]{CHSH}%
  \BibitemOpen
  \bibfield  {author} {\bibinfo {author} {\bibfnamefont {J.~F.}\ \bibnamefont
  {Clauser}}, \bibinfo {author} {\bibfnamefont {M.~A.}\ \bibnamefont {Horne}},
  \bibinfo {author} {\bibfnamefont {A.}~\bibnamefont {Shimony}}, \ and\
  \bibinfo {author} {\bibfnamefont {R.~A.}\ \bibnamefont {Holt}},\ }\href
  {\doibase 10.1103/PhysRevLett.23.880} {\bibfield  {journal} {\bibinfo
  {journal} {Phys. Rev. Lett.}\ }\textbf {\bibinfo {volume} {23}},\ \bibinfo
  {pages} {880} (\bibinfo {year} {1969})}\BibitemShut {NoStop}%
\bibitem [{\citenamefont {Cabello}(2001)}]{Cabello_2001}%
  \BibitemOpen
  \bibfield  {author} {\bibinfo {author} {\bibfnamefont {A.}~\bibnamefont
  {Cabello}},\ }\href {\doibase 10.1103/PhysRevLett.87.010403} {\bibfield
  {journal} {\bibinfo  {journal} {Phys. Rev. Lett.}\ }\textbf {\bibinfo
  {volume} {87}},\ \bibinfo {pages} {10403} (\bibinfo {year}
  {2001})}\BibitemShut {NoStop}%
\bibitem [{\citenamefont {Simon}\ and\ \citenamefont
  {Irvine}(2003)}]{Simon_Irvine_2003}%
  \BibitemOpen
  \bibfield  {author} {\bibinfo {author} {\bibfnamefont {C.}~\bibnamefont
  {Simon}}\ and\ \bibinfo {author} {\bibfnamefont {W.~T.~M.}\ \bibnamefont
  {Irvine}},\ }\href {\doibase 10.1103/PhysRevLett.91.110405} {\bibfield
  {journal} {\bibinfo  {journal} {Phys. Rev. Lett.}\ }\textbf {\bibinfo
  {volume} {91}},\ \bibinfo {pages} {110405} (\bibinfo {year}
  {2003})}\BibitemShut {NoStop}%
\bibitem [{\citenamefont {Matsukevich}\ \emph {et~al.}(2008)\citenamefont
  {Matsukevich}, \citenamefont {Maunz}, \citenamefont {Moehring}, \citenamefont
  {Olmschenk},\ and\ \citenamefont {Monroe}}]{Matsukevich_Monroe_2008}%
  \BibitemOpen
  \bibfield  {author} {\bibinfo {author} {\bibfnamefont {D.~N.}\ \bibnamefont
  {Matsukevich}}, \bibinfo {author} {\bibfnamefont {P.}~\bibnamefont {Maunz}},
  \bibinfo {author} {\bibfnamefont {D.~L.}\ \bibnamefont {Moehring}}, \bibinfo
  {author} {\bibfnamefont {S.}~\bibnamefont {Olmschenk}}, \ and\ \bibinfo
  {author} {\bibfnamefont {C.}~\bibnamefont {Monroe}},\ }\href {\doibase
  10.1103/PhysRevLett.100.150404} {\bibfield  {journal} {\bibinfo  {journal}
  {Phys. Rev. Lett.}\ }\textbf {\bibinfo {volume} {100}},\ \bibinfo {pages}
  {150404} (\bibinfo {year} {2008})}\BibitemShut {NoStop}%
\bibitem [{\citenamefont {Duan}\ \emph {et~al.}(2004)\citenamefont {Duan},
  \citenamefont {Blinov}, \citenamefont {Moehring},\ and\ \citenamefont
  {Monroe}}]{Duan_Monroe_2004}%
  \BibitemOpen
  \bibfield  {author} {\bibinfo {author} {\bibfnamefont {L.-M.}\ \bibnamefont
  {Duan}}, \bibinfo {author} {\bibfnamefont {B.~B.}\ \bibnamefont {Blinov}},
  \bibinfo {author} {\bibfnamefont {D.~L.}\ \bibnamefont {Moehring}}, \ and\
  \bibinfo {author} {\bibfnamefont {C.}~\bibnamefont {Monroe}},\ }\href
  {http://www.rintonpress.com/xqic4/qic-4-3/165-173.pdf} {\bibfield  {journal}
  {\bibinfo  {journal} {Quantum Information and Computation}\ }\textbf
  {\bibinfo {volume} {4}},\ \bibinfo {pages} {165} (\bibinfo {year}
  {2004})}\BibitemShut {NoStop}%
\bibitem [{\citenamefont {Jiang}\ \emph {et~al.}(2007)\citenamefont {Jiang},
  \citenamefont {Taylor}, \citenamefont {S\"orensen},\ and\ \citenamefont
  {Lukin}}]{Jiang_Lukin_2007}%
  \BibitemOpen
  \bibfield  {author} {\bibinfo {author} {\bibfnamefont {L.}~\bibnamefont
  {Jiang}}, \bibinfo {author} {\bibfnamefont {J.~M.}\ \bibnamefont {Taylor}},
  \bibinfo {author} {\bibfnamefont {A.~S.}\ \bibnamefont {S\"orensen}}, \ and\
  \bibinfo {author} {\bibfnamefont {M.~D.}\ \bibnamefont {Lukin}},\ }\href
  {\doibase 10.1103/PhysRevA.76.062323} {\bibfield  {journal} {\bibinfo
  {journal} {Phys. Rev. A}\ }\textbf {\bibinfo {volume} {76}},\ \bibinfo
  {pages} {62323} (\bibinfo {year} {2007})}\BibitemShut {NoStop}%
\bibitem [{\citenamefont {Devoret}\ and\ \citenamefont
  {Schoelkopf}(2013)}]{Devoret_Schoelkopf_2013}%
  \BibitemOpen
  \bibfield  {author} {\bibinfo {author} {\bibfnamefont {M.~H.}\ \bibnamefont
  {Devoret}}\ and\ \bibinfo {author} {\bibfnamefont {R.~J.}\ \bibnamefont
  {Schoelkopf}},\ }\href {\doibase 10.1126/science.1231930} {\bibfield
  {journal} {\bibinfo  {journal} {Science}\ }\textbf {\bibinfo {volume}
  {339}},\ \bibinfo {pages} {1169} (\bibinfo {year} {2013})}\BibitemShut
  {NoStop}%
\bibitem [{\citenamefont {Monroe}\ \emph {et~al.}(2014)\citenamefont {Monroe},
  \citenamefont {Raussendorf}, \citenamefont {Ruthven}, \citenamefont {Brown},
  \citenamefont {Maunz}, \citenamefont {Duan},\ and\ \citenamefont
  {Kim}}]{Monroe_Kim_2014}%
  \BibitemOpen
  \bibfield  {author} {\bibinfo {author} {\bibfnamefont {C.}~\bibnamefont
  {Monroe}}, \bibinfo {author} {\bibfnamefont {R.}~\bibnamefont {Raussendorf}},
  \bibinfo {author} {\bibfnamefont {A.}~\bibnamefont {Ruthven}}, \bibinfo
  {author} {\bibfnamefont {K.~R.}\ \bibnamefont {Brown}}, \bibinfo {author}
  {\bibfnamefont {P.}~\bibnamefont {Maunz}}, \bibinfo {author} {\bibfnamefont
  {L.-M.}\ \bibnamefont {Duan}}, \ and\ \bibinfo {author} {\bibfnamefont
  {J.}~\bibnamefont {Kim}},\ }\href {\doibase 10.1103/PhysRevA.89.022317}
  {\bibfield  {journal} {\bibinfo  {journal} {Phys. Rev. A}\ }\textbf {\bibinfo
  {volume} {89}},\ \bibinfo {pages} {22317} (\bibinfo {year}
  {2014})}\BibitemShut {NoStop}%
\bibitem [{\citenamefont {Cabrillo}\ \emph {et~al.}(1999)\citenamefont
  {Cabrillo}, \citenamefont {Cirac}, \citenamefont {Garcia-Fernandez},\ and\
  \citenamefont {Zoller}}]{Cabrillo_Zoller_1999}%
  \BibitemOpen
  \bibfield  {author} {\bibinfo {author} {\bibfnamefont {C.}~\bibnamefont
  {Cabrillo}}, \bibinfo {author} {\bibfnamefont {J.~I.}\ \bibnamefont {Cirac}},
  \bibinfo {author} {\bibfnamefont {P.}~\bibnamefont {Garcia-Fernandez}}, \
  and\ \bibinfo {author} {\bibfnamefont {P.}~\bibnamefont {Zoller}},\ }\href
  {\doibase 10.1103/PhysRevA.59.1025} {\bibfield  {journal} {\bibinfo
  {journal} {Phys. Rev. A}\ }\textbf {\bibinfo {volume} {59}},\ \bibinfo
  {pages} {1025} (\bibinfo {year} {1999})}\BibitemShut {NoStop}%
\bibitem [{\citenamefont {Duan}\ \emph {et~al.}(2001)\citenamefont {Duan},
  \citenamefont {Lukin}, \citenamefont {Cirac},\ and\ \citenamefont
  {Zoller}}]{DLCZ}%
  \BibitemOpen
  \bibfield  {author} {\bibinfo {author} {\bibfnamefont {L.-M.}\ \bibnamefont
  {Duan}}, \bibinfo {author} {\bibfnamefont {M.~D.}\ \bibnamefont {Lukin}},
  \bibinfo {author} {\bibfnamefont {J.~I.}\ \bibnamefont {Cirac}}, \ and\
  \bibinfo {author} {\bibfnamefont {P.}~\bibnamefont {Zoller}},\ }\href
  {\doibase 10.1038/35106500} {\bibfield  {journal} {\bibinfo  {journal}
  {Nature}\ }\textbf {\bibinfo {volume} {414}},\ \bibinfo {pages} {413}
  (\bibinfo {year} {2001})}\BibitemShut {NoStop}%
\bibitem [{\citenamefont {Barrett}\ and\ \citenamefont
  {Kok}(2004)}]{Barrett_Kok_2004}%
  \BibitemOpen
  \bibfield  {author} {\bibinfo {author} {\bibfnamefont {S.~D.}\ \bibnamefont
  {Barrett}}\ and\ \bibinfo {author} {\bibfnamefont {P.}~\bibnamefont {Kok}},\
  }\href {\doibase 10.1103/PhysRevA.71.060310} {\bibfield  {journal} {\bibinfo
  {journal} {Phys. Rev. A}\ }\textbf {\bibinfo {volume} {71}},\ \bibinfo
  {pages} {060310} (\bibinfo {year} {2004})}\BibitemShut {NoStop}%
\bibitem [{\citenamefont {Chou}\ \emph {et~al.}(2005)\citenamefont {Chou},
  \citenamefont {de~Riedmatten}, \citenamefont {Felinto}, \citenamefont
  {Polyakov}, \citenamefont {van Enk},\ and\ \citenamefont
  {Kimble}}]{Chou_Kimble_2005}%
  \BibitemOpen
  \bibfield  {author} {\bibinfo {author} {\bibfnamefont {C.~W.}\ \bibnamefont
  {Chou}}, \bibinfo {author} {\bibfnamefont {H.}~\bibnamefont {de~Riedmatten}},
  \bibinfo {author} {\bibfnamefont {D.}~\bibnamefont {Felinto}}, \bibinfo
  {author} {\bibfnamefont {S.~V.}\ \bibnamefont {Polyakov}}, \bibinfo {author}
  {\bibfnamefont {S.~J.}\ \bibnamefont {van Enk}}, \ and\ \bibinfo {author}
  {\bibfnamefont {H.~J.}\ \bibnamefont {Kimble}},\ }\href {\doibase
  10.1038/nature04353} {\bibfield  {journal} {\bibinfo  {journal} {Nature}\
  }\textbf {\bibinfo {volume} {438}},\ \bibinfo {pages} {828} (\bibinfo {year}
  {2005})}\BibitemShut {NoStop}%
\bibitem [{\citenamefont {Moehring}\ \emph {et~al.}(2007)\citenamefont
  {Moehring}, \citenamefont {Maunz}, \citenamefont {Olmschenk}, \citenamefont
  {Younge}, \citenamefont {Matsukevich}, \citenamefont {Duan},\ and\
  \citenamefont {Monroe}}]{Moehring_Monroe_2007}%
  \BibitemOpen
  \bibfield  {author} {\bibinfo {author} {\bibfnamefont {D.~L.}\ \bibnamefont
  {Moehring}}, \bibinfo {author} {\bibfnamefont {P.}~\bibnamefont {Maunz}},
  \bibinfo {author} {\bibfnamefont {S.}~\bibnamefont {Olmschenk}}, \bibinfo
  {author} {\bibfnamefont {K.~C.}\ \bibnamefont {Younge}}, \bibinfo {author}
  {\bibfnamefont {D.~N.}\ \bibnamefont {Matsukevich}}, \bibinfo {author}
  {\bibfnamefont {L.-M.}\ \bibnamefont {Duan}}, \ and\ \bibinfo {author}
  {\bibfnamefont {C.}~\bibnamefont {Monroe}},\ }\href {\doibase
  10.1038/nature06118} {\bibfield  {journal} {\bibinfo  {journal} {Nature}\
  }\textbf {\bibinfo {volume} {449}},\ \bibinfo {pages} {68} (\bibinfo {year}
  {2007})}\BibitemShut {NoStop}%
\bibitem [{\citenamefont {Hofmann}\ \emph {et~al.}(2012)\citenamefont
  {Hofmann}, \citenamefont {Krug}, \citenamefont {Ortegel}, \citenamefont
  {Gerard}, \citenamefont {Weber}, \citenamefont {Rosenfeld},\ and\
  \citenamefont {Weinfurter}}]{Hofmann_Weinfurter_2012}%
  \BibitemOpen
  \bibfield  {author} {\bibinfo {author} {\bibfnamefont {J.}~\bibnamefont
  {Hofmann}}, \bibinfo {author} {\bibfnamefont {M.}~\bibnamefont {Krug}},
  \bibinfo {author} {\bibfnamefont {N.}~\bibnamefont {Ortegel}}, \bibinfo
  {author} {\bibfnamefont {L.}~\bibnamefont {Gerard}}, \bibinfo {author}
  {\bibfnamefont {M.}~\bibnamefont {Weber}}, \bibinfo {author} {\bibfnamefont
  {W.}~\bibnamefont {Rosenfeld}}, \ and\ \bibinfo {author} {\bibfnamefont
  {H.}~\bibnamefont {Weinfurter}},\ }\href {\doibase 10.1126/science.1221856}
  {\bibfield  {journal} {\bibinfo  {journal} {Science}\ }\textbf {\bibinfo
  {volume} {337}},\ \bibinfo {pages} {72} (\bibinfo {year} {2012})}\BibitemShut
  {NoStop}%
\bibitem [{\citenamefont {Bernien}\ \emph {et~al.}(2013)\citenamefont
  {Bernien}, \citenamefont {Hensen}, \citenamefont {Pfaff}, \citenamefont
  {Koolstra}, \citenamefont {Blok}, \citenamefont {Robledo}, \citenamefont
  {Taminiau}, \citenamefont {Markham}, \citenamefont {Twitchen}, \citenamefont
  {Childress},\ and\ \citenamefont {Hanson}}]{Bernien_Hanson_2013}%
  \BibitemOpen
  \bibfield  {author} {\bibinfo {author} {\bibfnamefont {H.}~\bibnamefont
  {Bernien}}, \bibinfo {author} {\bibfnamefont {B.}~\bibnamefont {Hensen}},
  \bibinfo {author} {\bibfnamefont {W.}~\bibnamefont {Pfaff}}, \bibinfo
  {author} {\bibfnamefont {G.}~\bibnamefont {Koolstra}}, \bibinfo {author}
  {\bibfnamefont {M.~S.}\ \bibnamefont {Blok}}, \bibinfo {author}
  {\bibfnamefont {L.}~\bibnamefont {Robledo}}, \bibinfo {author} {\bibfnamefont
  {T.~H.}\ \bibnamefont {Taminiau}}, \bibinfo {author} {\bibfnamefont
  {M.}~\bibnamefont {Markham}}, \bibinfo {author} {\bibfnamefont {D.~J.}\
  \bibnamefont {Twitchen}}, \bibinfo {author} {\bibfnamefont {L.}~\bibnamefont
  {Childress}}, \ and\ \bibinfo {author} {\bibfnamefont {R.}~\bibnamefont
  {Hanson}},\ }\href {\doibase 10.1038/nature12016} {\bibfield  {journal}
  {\bibinfo  {journal} {Nature}\ }\textbf {\bibinfo {volume} {497}},\ \bibinfo
  {pages} {86} (\bibinfo {year} {2013})}\BibitemShut {NoStop}%
\bibitem [{\citenamefont {Roy}\ \emph {et~al.}(2015)\citenamefont {Roy},
  \citenamefont {Jiang}, \citenamefont {Stone},\ and\ \citenamefont
  {Devoret}}]{Roy_Devoret_2015}%
  \BibitemOpen
  \bibfield  {author} {\bibinfo {author} {\bibfnamefont {A.}~\bibnamefont
  {Roy}}, \bibinfo {author} {\bibfnamefont {L.}~\bibnamefont {Jiang}}, \bibinfo
  {author} {\bibfnamefont {A.~D.}\ \bibnamefont {Stone}}, \ and\ \bibinfo
  {author} {\bibfnamefont {M.}~\bibnamefont {Devoret}},\ }\href {\doibase
  10.1103/PhysRevLett.115.150503} {\bibfield  {journal} {\bibinfo  {journal}
  {Phys. Rev. Lett.}\ }\textbf {\bibinfo {volume} {115}},\ \bibinfo {pages}
  {150503} (\bibinfo {year} {2015})}\BibitemShut {NoStop}%
\bibitem [{\citenamefont {Silveri}\ \emph {et~al.}(2015)\citenamefont
  {Silveri}, \citenamefont {Zalys-Geller}, \citenamefont {Hatridge},
  \citenamefont {Leghtas}, \citenamefont {Devoret},\ and\ \citenamefont
  {Girvin}}]{Silveri_Girvin_2015}%
  \BibitemOpen
  \bibfield  {author} {\bibinfo {author} {\bibfnamefont {M.}~\bibnamefont
  {Silveri}}, \bibinfo {author} {\bibfnamefont {E.}~\bibnamefont
  {Zalys-Geller}}, \bibinfo {author} {\bibfnamefont {M.}~\bibnamefont
  {Hatridge}}, \bibinfo {author} {\bibfnamefont {Z.}~\bibnamefont {Leghtas}},
  \bibinfo {author} {\bibfnamefont {M.}~\bibnamefont {Devoret}}, \ and\
  \bibinfo {author} {\bibfnamefont {S.~M.}\ \bibnamefont {Girvin}},\ }\href
  {http://arxiv.org/abs/1507.00732} {\bibfield  {journal} {\bibinfo  {journal}
  {arXiv:1507.00732}\ } (\bibinfo {year} {2015})}\BibitemShut {NoStop}%
\bibitem [{\citenamefont {Jeong}\ and\ \citenamefont
  {Kim}(2002)}]{Jeong_Kim_2002}%
  \BibitemOpen
  \bibfield  {author} {\bibinfo {author} {\bibfnamefont {H.}~\bibnamefont
  {Jeong}}\ and\ \bibinfo {author} {\bibfnamefont {M.~S.}\ \bibnamefont
  {Kim}},\ }\href {\doibase 10.1103/PhysRevA.65.042305} {\bibfield  {journal}
  {\bibinfo  {journal} {Phys. Rev. A}\ }\textbf {\bibinfo {volume} {65}},\
  \bibinfo {pages} {42305} (\bibinfo {year} {2002})}\BibitemShut {NoStop}%
\bibitem [{\citenamefont {Ralph}\ \emph {et~al.}(2002)\citenamefont {Ralph},
  \citenamefont {Munro},\ and\ \citenamefont {Milburn}}]{Ralph_Milburn_2002}%
  \BibitemOpen
  \bibfield  {author} {\bibinfo {author} {\bibfnamefont {T.~C.}\ \bibnamefont
  {Ralph}}, \bibinfo {author} {\bibfnamefont {W.~J.}\ \bibnamefont {Munro}}, \
  and\ \bibinfo {author} {\bibfnamefont {G.~J.}\ \bibnamefont {Milburn}},\
  }\href {\doibase 10.1117/12.483016} {\bibfield  {journal} {\bibinfo
  {journal} {Proc. SPIE}\ }\textbf {\bibinfo {volume} {4917}},\ \bibinfo
  {pages} {1} (\bibinfo {year} {2002})}\BibitemShut {NoStop}%
\bibitem [{\citenamefont {Leghtas}\ \emph {et~al.}(2013)\citenamefont
  {Leghtas}, \citenamefont {Kirchmair}, \citenamefont {Vlastakis},
  \citenamefont {Schoelkopf}, \citenamefont {Devoret},\ and\ \citenamefont
  {Mirrahimi}}]{Leghtas_Mirrahimi_2013}%
  \BibitemOpen
  \bibfield  {author} {\bibinfo {author} {\bibfnamefont {Z.}~\bibnamefont
  {Leghtas}}, \bibinfo {author} {\bibfnamefont {G.}~\bibnamefont {Kirchmair}},
  \bibinfo {author} {\bibfnamefont {B.}~\bibnamefont {Vlastakis}}, \bibinfo
  {author} {\bibfnamefont {R.~J.}\ \bibnamefont {Schoelkopf}}, \bibinfo
  {author} {\bibfnamefont {M.~H.}\ \bibnamefont {Devoret}}, \ and\ \bibinfo
  {author} {\bibfnamefont {M.}~\bibnamefont {Mirrahimi}},\ }\href {\doibase
  10.1103/PhysRevLett.111.120501} {\bibfield  {journal} {\bibinfo  {journal}
  {Phys. Rev. Lett.}\ }\textbf {\bibinfo {volume} {111}},\ \bibinfo {pages}
  {120501} (\bibinfo {year} {2013})}\BibitemShut {NoStop}%
\bibitem [{\citenamefont {Mirrahimi}\ \emph {et~al.}(2014)\citenamefont
  {Mirrahimi}, \citenamefont {Leghtas}, \citenamefont {Albert}, \citenamefont
  {Touzard}, \citenamefont {Schoelkopf}, \citenamefont {Jiang},\ and\
  \citenamefont {Devoret}}]{Mirrahimi_Devoret_2014}%
  \BibitemOpen
  \bibfield  {author} {\bibinfo {author} {\bibfnamefont {M.}~\bibnamefont
  {Mirrahimi}}, \bibinfo {author} {\bibfnamefont {Z.}~\bibnamefont {Leghtas}},
  \bibinfo {author} {\bibfnamefont {V.~V.}\ \bibnamefont {Albert}}, \bibinfo
  {author} {\bibfnamefont {S.}~\bibnamefont {Touzard}}, \bibinfo {author}
  {\bibfnamefont {R.~J.}\ \bibnamefont {Schoelkopf}}, \bibinfo {author}
  {\bibfnamefont {L.}~\bibnamefont {Jiang}}, \ and\ \bibinfo {author}
  {\bibfnamefont {M.~H.}\ \bibnamefont {Devoret}},\ }\href
  {http://stacks.iop.org/1367-2630/16/i=4/a=045014} {\bibfield  {journal}
  {\bibinfo  {journal} {New Journal of Physics}\ }\textbf {\bibinfo {volume}
  {16}},\ \bibinfo {pages} {45014} (\bibinfo {year} {2014})}\BibitemShut
  {NoStop}%
\bibitem [{\citenamefont {Ofek}\ \emph {et~al.}(2016)\citenamefont {Ofek},
  \citenamefont {Petrenko}, \citenamefont {Heeres}, \citenamefont {Reinhold},
  \citenamefont {Leghtas}, \citenamefont {Vlastakis}, \citenamefont {Liu},
  \citenamefont {Frunzio}, \citenamefont {Girvin}, \citenamefont {Jiang},
  \citenamefont {Mirrahimi}, \citenamefont {Devoret},\ and\ \citenamefont
  {Schoelkopf}}]{Ofek_Schoelkopf_2016}%
  \BibitemOpen
  \bibfield  {author} {\bibinfo {author} {\bibfnamefont {N.}~\bibnamefont
  {Ofek}}, \bibinfo {author} {\bibfnamefont {A.}~\bibnamefont {Petrenko}},
  \bibinfo {author} {\bibfnamefont {R.}~\bibnamefont {Heeres}}, \bibinfo
  {author} {\bibfnamefont {P.}~\bibnamefont {Reinhold}}, \bibinfo {author}
  {\bibfnamefont {Z.}~\bibnamefont {Leghtas}}, \bibinfo {author} {\bibfnamefont
  {B.}~\bibnamefont {Vlastakis}}, \bibinfo {author} {\bibfnamefont
  {Y.}~\bibnamefont {Liu}}, \bibinfo {author} {\bibfnamefont {L.}~\bibnamefont
  {Frunzio}}, \bibinfo {author} {\bibfnamefont {S.~M.}\ \bibnamefont {Girvin}},
  \bibinfo {author} {\bibfnamefont {L.}~\bibnamefont {Jiang}}, \bibinfo
  {author} {\bibfnamefont {M.}~\bibnamefont {Mirrahimi}}, \bibinfo {author}
  {\bibfnamefont {M.~H.}\ \bibnamefont {Devoret}}, \ and\ \bibinfo {author}
  {\bibfnamefont {R.~J.}\ \bibnamefont {Schoelkopf}},\ }\href
  {http://arxiv.org/abs/1602.04768} {\  (\bibinfo {year} {2016})},\ \Eprint
  {http://arxiv.org/abs/1602.04768} {arXiv:1602.04768} \BibitemShut {NoStop}%
\bibitem [{\citenamefont {Nemoto}\ and\ \citenamefont
  {Munro}(2004)}]{Nemoto_Munro_2004}%
  \BibitemOpen
  \bibfield  {author} {\bibinfo {author} {\bibfnamefont {K.}~\bibnamefont
  {Nemoto}}\ and\ \bibinfo {author} {\bibfnamefont {W.~J.}\ \bibnamefont
  {Munro}},\ }\href {\doibase 10.1103/PhysRevLett.93.250502} {\bibfield
  {journal} {\bibinfo  {journal} {Phys. Rev. Lett.}\ }\textbf {\bibinfo
  {volume} {93}},\ \bibinfo {pages} {250502} (\bibinfo {year}
  {2004})}\BibitemShut {NoStop}%
\bibitem [{\citenamefont {Spiller}\ \emph {et~al.}(2006)\citenamefont
  {Spiller}, \citenamefont {Nemoto}, \citenamefont {Braunstein}, \citenamefont
  {Munro}, \citenamefont {{Van Loock}},\ and\ \citenamefont
  {Milburn}}]{Spiller_MIlburn_2006}%
  \BibitemOpen
  \bibfield  {author} {\bibinfo {author} {\bibfnamefont {T.~P.}\ \bibnamefont
  {Spiller}}, \bibinfo {author} {\bibfnamefont {K.}~\bibnamefont {Nemoto}},
  \bibinfo {author} {\bibfnamefont {S.~L.}\ \bibnamefont {Braunstein}},
  \bibinfo {author} {\bibfnamefont {W.~J.}\ \bibnamefont {Munro}}, \bibinfo
  {author} {\bibfnamefont {P.}~\bibnamefont {{Van Loock}}}, \ and\ \bibinfo
  {author} {\bibfnamefont {G.~J.}\ \bibnamefont {Milburn}},\ }\href@noop {}
  {\bibfield  {journal} {\bibinfo  {journal} {New Journal of Physics}\ }\textbf
  {\bibinfo {volume} {8}} (\bibinfo {year} {2006})}\BibitemShut {NoStop}%
\bibitem [{\citenamefont {Roch}\ \emph {et~al.}(2014)\citenamefont {Roch},
  \citenamefont {Schwartz}, \citenamefont {Motzoi}, \citenamefont {Macklin},
  \citenamefont {Vijay}, \citenamefont {Eddins}, \citenamefont {Korotkov},
  \citenamefont {Whaley}, \citenamefont {Sarovar},\ and\ \citenamefont
  {Siddiqi}}]{Roch_Siddiqi_2014}%
  \BibitemOpen
  \bibfield  {author} {\bibinfo {author} {\bibfnamefont {N.}~\bibnamefont
  {Roch}}, \bibinfo {author} {\bibfnamefont {M.~E.}\ \bibnamefont {Schwartz}},
  \bibinfo {author} {\bibfnamefont {F.}~\bibnamefont {Motzoi}}, \bibinfo
  {author} {\bibfnamefont {C.}~\bibnamefont {Macklin}}, \bibinfo {author}
  {\bibfnamefont {R.}~\bibnamefont {Vijay}}, \bibinfo {author} {\bibfnamefont
  {A.~W.}\ \bibnamefont {Eddins}}, \bibinfo {author} {\bibfnamefont {A.~N.}\
  \bibnamefont {Korotkov}}, \bibinfo {author} {\bibfnamefont {K.~B.}\
  \bibnamefont {Whaley}}, \bibinfo {author} {\bibfnamefont {M.}~\bibnamefont
  {Sarovar}}, \ and\ \bibinfo {author} {\bibfnamefont {I.}~\bibnamefont
  {Siddiqi}},\ }\href {\doibase 10.1103/PhysRevLett.112.170501} {\bibfield
  {journal} {\bibinfo  {journal} {Phys. Rev. Lett.}\ }\textbf {\bibinfo
  {volume} {112}},\ \bibinfo {pages} {170501} (\bibinfo {year}
  {2014})}\BibitemShut {NoStop}%
\bibitem [{\citenamefont {Sarlette}\ and\ \citenamefont
  {Mirrahimi}(2016)}]{Sarlette_Mirrahimi_2016}%
  \BibitemOpen
  \bibfield  {author} {\bibinfo {author} {\bibfnamefont {A.}~\bibnamefont
  {Sarlette}}\ and\ \bibinfo {author} {\bibfnamefont {M.}~\bibnamefont
  {Mirrahimi}},\ }\href {http://arxiv.org/abs/1604.04490} {\  (\bibinfo {year}
  {2016})},\ \Eprint {http://arxiv.org/abs/1604.04490} {arXiv:1604.04490}
  \BibitemShut {NoStop}%
\bibitem [{\citenamefont {Wang}\ \emph {et~al.}(2016)\citenamefont {Wang},
  \citenamefont {Gao}, \citenamefont {Reinhold}, \citenamefont {Heeres},
  \citenamefont {Ofek}, \citenamefont {Chou}, \citenamefont {Axline},
  \citenamefont {Reagor}, \citenamefont {Blumoff}, \citenamefont {Sliwa},
  \citenamefont {Frunzio}, \citenamefont {Girvin}, \citenamefont {Jiang},
  \citenamefont {Mirrahimi}, \citenamefont {Devoret},\ and\ \citenamefont
  {Schoelkopf}}]{Wang_Schoelkopf_2016}%
  \BibitemOpen
  \bibfield  {author} {\bibinfo {author} {\bibfnamefont {C.}~\bibnamefont
  {Wang}}, \bibinfo {author} {\bibfnamefont {Y.~Y.}\ \bibnamefont {Gao}},
  \bibinfo {author} {\bibfnamefont {P.}~\bibnamefont {Reinhold}}, \bibinfo
  {author} {\bibfnamefont {R.~W.}\ \bibnamefont {Heeres}}, \bibinfo {author}
  {\bibfnamefont {N.}~\bibnamefont {Ofek}}, \bibinfo {author} {\bibfnamefont
  {K.}~\bibnamefont {Chou}}, \bibinfo {author} {\bibfnamefont {C.}~\bibnamefont
  {Axline}}, \bibinfo {author} {\bibfnamefont {M.}~\bibnamefont {Reagor}},
  \bibinfo {author} {\bibfnamefont {J.}~\bibnamefont {Blumoff}}, \bibinfo
  {author} {\bibfnamefont {K.~M.}\ \bibnamefont {Sliwa}}, \bibinfo {author}
  {\bibfnamefont {L.}~\bibnamefont {Frunzio}}, \bibinfo {author} {\bibfnamefont
  {S.~M.}\ \bibnamefont {Girvin}}, \bibinfo {author} {\bibfnamefont
  {L.}~\bibnamefont {Jiang}}, \bibinfo {author} {\bibfnamefont
  {M.}~\bibnamefont {Mirrahimi}}, \bibinfo {author} {\bibfnamefont {M.~H.}\
  \bibnamefont {Devoret}}, \ and\ \bibinfo {author} {\bibfnamefont {R.~J.}\
  \bibnamefont {Schoelkopf}},\ }\href
  {http://science.sciencemag.org/content/352/6289/1087.abstract} {\bibfield
  {journal} {\bibinfo  {journal} {Science}\ }\textbf {\bibinfo {volume}
  {352}},\ \bibinfo {pages} {1087} (\bibinfo {year} {2016})}\BibitemShut
  {NoStop}%
\bibitem [{\citenamefont {Krastanov}\ \emph {et~al.}(2015)\citenamefont
  {Krastanov}, \citenamefont {Albert}, \citenamefont {Shen}, \citenamefont
  {Zou}, \citenamefont {Heeres}, \citenamefont {Vlastakis}, \citenamefont
  {Schoelkopf},\ and\ \citenamefont {Jiang}}]{Krastanov_Jiang_2015}%
  \BibitemOpen
  \bibfield  {author} {\bibinfo {author} {\bibfnamefont {S.}~\bibnamefont
  {Krastanov}}, \bibinfo {author} {\bibfnamefont {V.~V.}\ \bibnamefont
  {Albert}}, \bibinfo {author} {\bibfnamefont {C.}~\bibnamefont {Shen}},
  \bibinfo {author} {\bibfnamefont {C.~L.}\ \bibnamefont {Zou}}, \bibinfo
  {author} {\bibfnamefont {R.~W.}\ \bibnamefont {Heeres}}, \bibinfo {author}
  {\bibfnamefont {B.}~\bibnamefont {Vlastakis}}, \bibinfo {author}
  {\bibfnamefont {R.~J.}\ \bibnamefont {Schoelkopf}}, \ and\ \bibinfo {author}
  {\bibfnamefont {L.}~\bibnamefont {Jiang}},\ }\href {\doibase
  10.1103/PhysRevA.92.040303} {\bibfield  {journal} {\bibinfo  {journal}
  {Physical Review A - Atomic, Molecular, and Optical Physics}\ }\textbf
  {\bibinfo {volume} {92}},\ \bibinfo {pages} {1} (\bibinfo {year}
  {2015})}\BibitemShut {NoStop}%
\bibitem [{\citenamefont {Heeres}\ \emph {et~al.}(2015)\citenamefont {Heeres},
  \citenamefont {Vlastakis}, \citenamefont {Holland}, \citenamefont
  {Krastanov}, \citenamefont {Albert}, \citenamefont {Frunzio}, \citenamefont
  {Jiang},\ and\ \citenamefont {Schoelkopf}}]{Heeres_Schoelkopf_2015}%
  \BibitemOpen
  \bibfield  {author} {\bibinfo {author} {\bibfnamefont {R.~W.}\ \bibnamefont
  {Heeres}}, \bibinfo {author} {\bibfnamefont {B.}~\bibnamefont {Vlastakis}},
  \bibinfo {author} {\bibfnamefont {E.}~\bibnamefont {Holland}}, \bibinfo
  {author} {\bibfnamefont {S.}~\bibnamefont {Krastanov}}, \bibinfo {author}
  {\bibfnamefont {V.~V.}\ \bibnamefont {Albert}}, \bibinfo {author}
  {\bibfnamefont {L.}~\bibnamefont {Frunzio}}, \bibinfo {author} {\bibfnamefont
  {L.}~\bibnamefont {Jiang}}, \ and\ \bibinfo {author} {\bibfnamefont {R.~J.}\
  \bibnamefont {Schoelkopf}},\ }\href {\doibase 10.1103/PhysRevLett.115.137002}
  {\bibfield  {journal} {\bibinfo  {journal} {Phys. Rev. Lett.}\ }\textbf
  {\bibinfo {volume} {115}},\ \bibinfo {pages} {137002} (\bibinfo {year}
  {2015})}\BibitemShut {NoStop}%
\bibitem [{\citenamefont {Cirac}\ \emph {et~al.}(1997)\citenamefont {Cirac},
  \citenamefont {Zoller}, \citenamefont {Kimble},\ and\ \citenamefont
  {Mabuchi}}]{Cirac_Mabuchi_1997}%
  \BibitemOpen
  \bibfield  {author} {\bibinfo {author} {\bibfnamefont {J.~I.}\ \bibnamefont
  {Cirac}}, \bibinfo {author} {\bibfnamefont {P.}~\bibnamefont {Zoller}},
  \bibinfo {author} {\bibfnamefont {H.~J.}\ \bibnamefont {Kimble}}, \ and\
  \bibinfo {author} {\bibfnamefont {H.}~\bibnamefont {Mabuchi}},\ }\href
  {\doibase 10.1103/PhysRevLett.78.3221} {\bibfield  {journal} {\bibinfo
  {journal} {Physical Review Letters}\ }\textbf {\bibinfo {volume} {78}},\
  \bibinfo {pages} {3221} (\bibinfo {year} {1997})}\BibitemShut {NoStop}%
\bibitem [{\citenamefont {Yin}\ \emph {et~al.}(2013)\citenamefont {Yin},
  \citenamefont {Chen}, \citenamefont {Sank}, \citenamefont {O'Malley},
  \citenamefont {White}, \citenamefont {Barends}, \citenamefont {Kelly},
  \citenamefont {Lucero}, \citenamefont {Mariantoni}, \citenamefont {Megrant},
  \citenamefont {Neill}, \citenamefont {Vainsencher}, \citenamefont {Wenner},
  \citenamefont {Korotkov}, \citenamefont {Cleland},\ and\ \citenamefont
  {Martinis}}]{Yin_Martinis_2013}%
  \BibitemOpen
  \bibfield  {author} {\bibinfo {author} {\bibfnamefont {Y.}~\bibnamefont
  {Yin}}, \bibinfo {author} {\bibfnamefont {Y.}~\bibnamefont {Chen}}, \bibinfo
  {author} {\bibfnamefont {D.}~\bibnamefont {Sank}}, \bibinfo {author}
  {\bibfnamefont {P.~J.~J.}\ \bibnamefont {O'Malley}}, \bibinfo {author}
  {\bibfnamefont {T.~C.}\ \bibnamefont {White}}, \bibinfo {author}
  {\bibfnamefont {R.}~\bibnamefont {Barends}}, \bibinfo {author} {\bibfnamefont
  {J.}~\bibnamefont {Kelly}}, \bibinfo {author} {\bibfnamefont
  {E.}~\bibnamefont {Lucero}}, \bibinfo {author} {\bibfnamefont
  {M.}~\bibnamefont {Mariantoni}}, \bibinfo {author} {\bibfnamefont
  {A.}~\bibnamefont {Megrant}}, \bibinfo {author} {\bibfnamefont
  {C.}~\bibnamefont {Neill}}, \bibinfo {author} {\bibfnamefont
  {A.}~\bibnamefont {Vainsencher}}, \bibinfo {author} {\bibfnamefont
  {J.}~\bibnamefont {Wenner}}, \bibinfo {author} {\bibfnamefont {A.~N.}\
  \bibnamefont {Korotkov}}, \bibinfo {author} {\bibfnamefont {A.~N.}\
  \bibnamefont {Cleland}}, \ and\ \bibinfo {author} {\bibfnamefont {J.~M.}\
  \bibnamefont {Martinis}},\ }\href {\doibase 10.1103/PhysRevLett.110.107001}
  {\bibfield  {journal} {\bibinfo  {journal} {Physical Review Letters}\
  }\textbf {\bibinfo {volume} {110}},\ \bibinfo {pages} {1} (\bibinfo {year}
  {2013})}\BibitemShut {NoStop}%
\bibitem [{\citenamefont {Leonhardt}\ and\ \citenamefont
  {Paul}(1995)}]{Leonhardt_Paul_1995}%
  \BibitemOpen
  \bibfield  {author} {\bibinfo {author} {\bibfnamefont {U.}~\bibnamefont
  {Leonhardt}}\ and\ \bibinfo {author} {\bibfnamefont {H.}~\bibnamefont
  {Paul}},\ }\href {\doibase http://dx.doi.org/10.1016/0079-6727(94)00007-L}
  {\bibfield  {journal} {\bibinfo  {journal} {Progress in Quantum Electronics}\
  }\textbf {\bibinfo {volume} {19}},\ \bibinfo {pages} {89} (\bibinfo {year}
  {1995})}\BibitemShut {NoStop}%
\bibitem [{\citenamefont {Steck}(2015)}]{Steck_2015}%
  \BibitemOpen
  \bibfield  {author} {\bibinfo {author} {\bibfnamefont {D.~A.}\ \bibnamefont
  {Steck}},\ }\href {http://steck.us/teaching} {\emph {\bibinfo {title}
  {{Quantum and Atom Optics}}}}\ (\bibinfo {year} {2015})\BibitemShut {NoStop}%
\bibitem [{\citenamefont {Gottesman}\ \emph {et~al.}(2001)\citenamefont
  {Gottesman}, \citenamefont {Kitaev},\ and\ \citenamefont
  {Preskill}}]{Gottesman_Preskill_2001}%
  \BibitemOpen
  \bibfield  {author} {\bibinfo {author} {\bibfnamefont {D.}~\bibnamefont
  {Gottesman}}, \bibinfo {author} {\bibfnamefont {A.}~\bibnamefont {Kitaev}}, \
  and\ \bibinfo {author} {\bibfnamefont {J.}~\bibnamefont {Preskill}},\ }\href
  {\doibase 10.1103/PhysRevA.64.012310} {\bibfield  {journal} {\bibinfo
  {journal} {Phys. Rev. A}\ }\textbf {\bibinfo {volume} {64}},\ \bibinfo
  {pages} {12310} (\bibinfo {year} {2001})}\BibitemShut {NoStop}%
\bibitem [{\citenamefont {Michael}\ \emph {et~al.}(2016)\citenamefont
  {Michael}, \citenamefont {Silveri}, \citenamefont {Brierley}, \citenamefont
  {Albert}, \citenamefont {Salmilehto}, \citenamefont {Jiang},\ and\
  \citenamefont {Girvin}}]{Michael_Girvin_2016}%
  \BibitemOpen
  \bibfield  {author} {\bibinfo {author} {\bibfnamefont {M.~H.}\ \bibnamefont
  {Michael}}, \bibinfo {author} {\bibfnamefont {M.}~\bibnamefont {Silveri}},
  \bibinfo {author} {\bibfnamefont {R.~T.}\ \bibnamefont {Brierley}}, \bibinfo
  {author} {\bibfnamefont {V.~V.}\ \bibnamefont {Albert}}, \bibinfo {author}
  {\bibfnamefont {J.}~\bibnamefont {Salmilehto}}, \bibinfo {author}
  {\bibfnamefont {L.}~\bibnamefont {Jiang}}, \ and\ \bibinfo {author}
  {\bibfnamefont {S.~M.}\ \bibnamefont {Girvin}},\ }\href
  {http://arxiv.org/abs/1602.00008} {\  (\bibinfo {year} {2016})},\ \Eprint
  {http://arxiv.org/abs/1602.00008} {arXiv:1602.00008} \BibitemShut {NoStop}%
\bibitem [{\citenamefont {Terhal}\ and\ \citenamefont
  {Weigand}(2016)}]{Terhal_Weigand_2016}%
  \BibitemOpen
  \bibfield  {author} {\bibinfo {author} {\bibfnamefont {B.~M.}\ \bibnamefont
  {Terhal}}\ and\ \bibinfo {author} {\bibfnamefont {D.}~\bibnamefont
  {Weigand}},\ }\href {\doibase 10.1103/PhysRevA.93.012315} {\bibfield
  {journal} {\bibinfo  {journal} {Physical Review A}\ }\textbf {\bibinfo
  {volume} {93}},\ \bibinfo {pages} {1} (\bibinfo {year} {2016})}\BibitemShut
  {NoStop}%
\end{thebibliography}%

\end{document}